\documentclass[12pt,notitlepage,superscriptaddress,amsmath,amssymb,aps,
showkeys]{revtex4-1}

\usepackage{dcolumn}
\usepackage{bm}
\usepackage{hyperref}
\usepackage{physics}
\usepackage{mathtools}
\usepackage{subcaption}
\usepackage{mathrsfs}
\allowdisplaybreaks

\def\p{\partial}

\usepackage{graphicx,amssymb,amsmath,amsthm,amsfonts}
\usepackage[usenames]{color}
\usepackage{xcolor}

\makeatletter
\newcommand\emailx[1]{%
\move@AF%
\def\@affil{{\normalfont\,#1\strut}{}}
}
\newcommand{\sVert}[1][0]{
  \ifcase#1\relax
  \rvert\or\bigr|\or\Bigr|\or\biggr|\or\Biggr
  \fi
}

\begin{document}

\title{
Canonical ensemble of a self-gravitating matter thin shell\\ in
AdS space}


\author{Tiago V. Fernandes}
\email{tiago.vasques.fernandes@tecnico.ulisboa.pt}
\affiliation{Center for Astrophysics and Gravitation -
CENTRA, Departamento de F\'{\i}sica,
Instituto Superior T\'{e}cnico - IST, Universidade de Lisboa - UL,
Avenida Rovisco Pais 1, 1049-001, Portugal.}

\author{Francisco J. Gandum}
\email{francisco.gandum@tecnico.ulisboa.pt}
\affiliation{Center for Astrophysics and Gravitation -
CENTRA, Departamento de F\'{\i}sica,
Instituto Superior T\'{e}cnico - IST, Universidade de Lisboa - UL,
Avenida Rovisco Pais 1, 1049-001, Portugal.} 

\author{Jos\'{e} P. S. Lemos}
\email{joselemos@ist.utl.pt}
\affiliation{Center for Astrophysics and Gravitation -
CENTRA, Departamento de F\'{\i}sica,
Instituto Superior T\'{e}cnico - IST, Universidade de Lisboa - UL,
Avenida Rovisco Pais 1, 1049-001, Portugal.} 

\begin{abstract}

We build the canonical ensemble of a hot self-gravitating matter thin
shell in anti-de Sitter (AdS) space by finding its partition function
through the Euclidean path integral approach with fixed temperature at
the conformal boundary. We obtain the reduced action of the system by
restricting the path integral to spherically symmetric metrics with
given boundary conditions and with the Hamiltonian constraint
satisfied.  The stationary conditions, i.e., the mechanical
equilibrium and the thermodynamic equilibrium, are obtained from
minimizing the reduced action.  Evaluating the perturbed reduced
action at the stationary points yields the mechanical stability
condition and the thermodynamic stability condition.
The reduced action calculated at the stationary points gives the
partition function in the zero-loop approximation and from it the
thermodynamic properties of the system are acquired. Within
thermodynamics alone, the only stability condition that one can
establish is thermodynamic stability, which follows from the
computation of the heat capacity.
For given specific pressure and temperature
equations of state for the shell, we obtain the solutions of
the ensemble.  There are four different thin shell solutions, one of
them is fully stable, i.e., is stable mechanically and
thermodynamically.  For the equations of state given, we find a first
order phase transition from the matter thermodynamic phase to the
Hawking-Page black hole phase. Moreover, there is a maximum
temperature above which the shell ceases to exist, presumably
at these high temperatures the shell inevitably collapses to
a black hole.

\end{abstract}

\keywords{Statistical ensembles, matter thin shell, anti-de Sitter, 
black hole thermodynamics}
\vspace{0.5cm}

\maketitle

\section{Introduction}

Lumps of excited matter fields have temperature. 
Lumps of excited gravitational fields also have temperature, 
the most notable case
being a black hole. The interface of contact between a black hole and
the external world is the event horizon, and so it is expected that
this horizon characterizes the black hole temperature itself.
The sequence of developments that led to the realization that 
black holes are objects with temperature and have thermodynamic
properties was gradual and intriguing.
It started with the possibility of energy extraction from a black
hole and its consequences, notably
the appearance of the 
irreducible mass concept and the  demonstration that
the black hole event horizon area always increases.
This indicated that, possibly, fundamental phenomena were lurking behind 
what could be expected.
These anticipations were confirmed when Bekenstein
\cite{Bekenstein:1973}, in order to save the second law of
thermodynamics, advanced the hypothesis that black holes possess
entropy proportional to its area,
paving the way to envisage black holes in a new light, namely,
as thermodynamic objects.
The black hole thermodynamic characterization was pushed forward when it
was found that there exist four laws of black hole mechanics
\cite{bardeen:1973} in striking analogy to the four laws of
thermodynamics.
It was then finalized by Hawking~\cite{hawking:1975} with the discovery
that black holes interacting with quantum fields emit radiation
with a black body spectrum, at the
Hawking temperature $T_{\rm H} = \frac{\kappa}{2\pi}$, where $\kappa$
is the surface gravity of the horizon. With this expression for the
temperature and the first law of black hole mechanics, which is the
energy conservation law, the entropy could be determined exactly with
the factor of one over four, i.e., a black hole possessed the
Bekenstein-Hawking entropy given by $S= \frac14\frac{A_+}{l_{\rm p}^2}$,
where $A_+$ is the area of the event horizon and $l_{\rm p}$ is the Planck
length.
The connection between black holes, thermodynamics, and quantum fields
put black holes in the center of physics phenomena. It led to an
intensive research on quantum fields in curved spacetimes, on the
quantum nature of spacetime, and to the study of quantum statistical
mechanics through the construction of statistical ensembles in curved
spacetimes.

In relation to statistical ensembles, the development came from the
Euclidean path integral approach of Feynman applied to gravitational
systems. It is known that, when Euclideanized, the path integral gives
statistical mechanics, so when the same procedure is made to black
hole spacetimes, the path integral should give statistical mechanics
of black holes.  The idea is that one can calculate the partition
function of stationary black hole
spaces through the Euclidean path integral approach to quantum
gravity, i.e., the partition function is formally given by $Z = \int
Dg_{\alpha \beta}\, {\rm e}^{-I[g_{\mu\nu}]}$, 
where $g_{\alpha \beta}$ is the Euclidean metric 
and $I$ is the Euclidean action.
The path integral in curved spaces possesses a number of difficulties
which are still unresolved, but at some level one can proceed.
Gibbons and Hawking~\cite{2gibbonshawking:1977} used the Euclidean
Schwarzschild solution to find the corresponding action and study the
canonical ensemble of the system, where the temperature is fixed at
spatial infinity. In order to deal with the path integral, 
the saddle point approximation was performed,
where the action would be 
expanded around the Euclidean Einstein solutions and only the 
zeroth order term would be dealt with, justifying the usage of the
action evaluated at the 
classical solutions to describe the partition function of the canonical 
ensemble.
They further used the Euclidean
Reissner-Nordstr\"om and Kerr black hole solutions to find the
corresponding actions in the zero order approximation,
and study the grand canonical ensembles of those
systems, where the temperature and the appropriate potential are fixed
at spatial infinity. Yet, for these black holes, and
in particular, for the Schwarzschild black hole, it was
found that the heat capacity was negative, indicating thermodynamic
instability and showing that, under small thermal perturbations, 
these black holes either evaporate or grow indefinitely~\cite{hawking:1976}.
Taking into account up to second order terms in the
expansion of the action 
named as one-loop approximation, which amounts to performing
quantum perturbations of the background metric, it was found that such
unstable Schwarzschild solutions had an unstable mode of
perturbations that allowed the tunneling of hot flat space to a
black hole state which could then grow
indefinitely~\cite{grossperryyaffe:1982}, an indication that hot flat
space was not a global minimum of the action. On the other hand, if
one examines a Schwarzschild black hole in anti-de Sitter
space, i.e., AdS space, there is still a black
hole solution analogue to the Gibbons-Hawking unstable solution but
another solution arises which is thermodynamically stable, and
additionally, hot AdS space can undergo a first order phase transition
to the stable black hole~\cite{hawkingpage1983}.
By enclosing the Schwarzschild black hole
in a spherical box with fixed radius and fixed temperature at its
boundary, it was found that if the box is small enough then the
unstable mode in Schwarzschild geometry ceases to exist
\cite{Allen:1984}.
Since AdS is a natural box, and an assembled box acts as a stabilizer
of the geometry, this motivated York to enclose the pure Schwarzschild
black hole inside a finite cavity to study the corresponding canonical
ensemble~\cite{York:1986}. York used also the classical solutions of the 
ensemble to compute the action and so the free energy of the system. 
However, to analyze the stability and phase transitions, he built a 
generalized free energy where the relation between the temperature 
and the horizon radius was relaxed. The York formalism was then formally 
constructed in~\cite{whitingyork:1988}, where the path integral was 
constrained to the hypersurface of paths obeying the Hamiltonian 
and momentum constraints, yielding a reduced action corresponding to the 
generalized free energy of the spherically symmetric black hole inside 
a cavity. One can then perform the 
saddle point approximation by expanding the reduced action 
around its stationary points and one can obtain the
expansion up to second order 
corrections to analyze stability in the hypersurface of constrained 
paths. Now, if one constructs
only the contributions of the stationary points, one is performing the
zero-loop approximation, which is then equivalent to the
Gibbons-Hawking approach. For the black hole in a cavity, it was
found that two black hole solutions also exist, with the smallest
being unstable and the largest being stable, and also that hot flat
space could undergo a first order phase transition to the stable black
hole. Thus, the York formalism makes the canonical ensemble of
asymptotically flat black holes valid, while the canonical ensemble of
AdS black holes is valid naturally. Within this
formalism one can even obtain
quantum gravity
corrections by calculating
the one-loop terms of the constrained path integral.
Further developments were made in~\cite{martinezyork:1989,
zaslavskii:1990,bradenbrownwhitingyork1990,browncomer:1994,
Lemos:1996,pecalemos1999,pecalemos2000,
Akbar:2010, luroyxiao:2010, andrelemos:2020, andrelemos:2021,
fernandeslemos2023,
lemoszaslavskii2023,lemoszaslavskii:2024,fernandeslemos2024} for
several different types of black holes in
spaces with negative, zero, or positive cosmological constant.

It is also important to establish in a self-consistent manner
the ways that gravity and matter
interact and the effects of this interplay on the
thermodynamics of gravitating matter systems.
An effective useful model is the self-gravitating matter thin
shell. In~\cite{martinez1996}, the first law of thermodynamics was
applied to these shells, and it was found that the
Einstein equations give an equation of state for the pressure and that
the temperature equation of state has a precise dependence on the
gravitational radius of the system.  The case of electrically charged
shells was first treated in~\cite{lemosquinta2015}. The results
were extended to arbitrary dimensions
in~\cite{andrelemosquinta2019,fernandeslemos2022}. A feature of such a
shell is that one can pick the same
temperature equation of state of the black
hole for the shell and put it very close to its gravitational radius,
one then has an object mimicking the thermodynamics of a black hole.
For shells made of extremal matter, this can be done smoothly, while
for the other cases, issues of stability and diverging pressure may
arise.
Another instance where the first law of thermodynamics of
self-gravitating systems can be used with benefit is in the quasiblack
hole approach, where matter systems on the verge of becoming a black
hole are analyzed \cite{Lemos:2009,Lemos:2010,Lemos:2020}.

In this work, we analyze a more fundamental way of studying the
thermodynamics of self-gravitating matter systems by building their
statistical ensemble.
This was sketched first in~\cite{martinezyork:1989}, where a thin
shell was considered with a black hole inside, and it was found that
the entropy of the system is the sum of the entropies of the shell and
the black hole. This was further worked in detail
in~\cite{lemoszaslavskii2023}, where it was shown that the entropy and
the thermodynamics are given simply as a function of the gravitational
radius of the system.
We develop these ideas and explore the matter sector of spherically
symmetric AdS spaces. For that, we carefully build the canonical
ensemble of a hot self-gravitating matter thin shell in AdS
using York's path integral formalism.
By
establishing the reduced action through the constraint equations, we
are able to show that the path integral depends only on the
gravitational radius of the system and on the radius of the shell.
One stationary point gives the mechanical solution
which shows that the radius of the shell depends only on the
gravitational radius, making the action depending solely on this latter
radius.  The other stationary point yields the thermodynamic solution
which sets the gravitational radius as a function of the temperature
of the ensemble. The zero order approximation then gives the partition
function as a function of this temperature.
Through perturbations of the ensemble, we analyze the corresponding
stability criteria of the system.  The solutions for the stationary
points of the action depend on the choice of the equation of state of
the matter. We give a linear equation of state for the pressure and an
equation of state for the temperature that allow for stable
solutions. From the zero-loop approximation, we obtain the
thermodynamics of the system, namely the free energy, the mean energy,
the entropy, and the heat capacity.
We compare the free energy of the self-gravitating matter thin shell
with the free energy of the black hole sector and analyze the
existence of first order phase transitions in AdS. In this way, we
simulate the interaction of matter and gravitation both treated at
zero-loop level with striking results.
One can include results at one-loop approximation valid for
nonself-gravitating radiation in AdS space and compare with the results
at zero-loop approximation found for the self-gravitating hot
shell. Our analysis is valid for microscopic configurations not near
the Planck scale.

The paper is organized as follows. 
In Sec.~\ref{sec:pathintegralconstrained}, we work out the canonical
ensemble, characterized by a fixed temperature, of a hot
self-gravitating
matter thin shell in AdS space. We define the partition function of
the system through the Euclidean path integral approach, postulate the
general Euclidean action and the matter free energy, impose spherical
symmetry and provide the boundary conditions.  We then enforce the
Hamiltonian and momentum constraints to finally find the reduced
action. We show that at this stage the partition function reduces to a
sum over metrics with different gravitational radii and different
radii of the shell.
In Sec.~\ref{sec:Zeroloopapproximation}, we perform the zero-loop
approximation, i.e., we investigate the reduced action evaluated at its
stationary points. The mechanical stationary point
yields that the radius of the shell is a function of
its gravitational radius and so
the partition function reduces to a
sum over metrics with different gravitational radii alone.
The thermodynamic stationary point
yields that the gravitational radius is a function of the
ensemble temperature, and so
finally, we find that
the partition function is a function of the temperature
alone, which is fixed.
We also find the stability conditions, and study
mechanical and thermodynamic stability of the shell.
In Sec.~\ref{sec:thermodynamics},
we derive the thermodynamics from the path integral in the
zero-loop approximation and show that
positive heat capacity for the system is equivalent to
the ensemble thermodynamic stability.
In Sec.~\ref{sec:unspecifiedeos}, we give specific equations of state
for the matter, namely, the matter pressure equation
and the matter temperature equation to
fully
determine the system. We find the radius of the shell as a function
of its gravitational radius,
the matter entropy as a
function of the gravitational radius, and
the gravitational radius as a
function of the ensemble temperature. In Sec.~\ref{tsxbh}, we compare
the results for the hot thin shell in AdS with the results for the
Hawking-Page black hole in AdS, notably we find the favorable states
for different ensemble temperatures.
In Sec.~\ref{sec:concl}, we conclude.

We choose units where the Planck constant is $\hbar = 1$, the
Boltzmann constant is $k=1$ and the
speed of light  is $c=1$. The
gravitational constant is written as the square of the
Planck length
$G= l_{\rm p}^2$ in
these units, while mass is understood as the inverse of the reduced
Compton wavelength associated to that mass, and temperature is
understood as the inverse
of a thermal wavelength. We adopt Greek indices for tensors
defined on the whole space and Latin indices for
tensors defined on hypersurfaces of constant spacelike coordinates.

\vfill

\section{Canonical ensemble of a hot
self-gravitating matter thin shell in
asymptotically AdS space and the reduced action
\label{sec:pathintegralconstrained}}

\subsection{The partition function given by the Euclidean
path integral approach}

Generally, in quantum systems represented by fields $\psi$
at temperature $T$, one can
build the statistical mechanics partition function $Z$
with expression 
$Z = \Tr({\rm e}^{-\beta H})$,
where $\beta=\frac{1}{T}$ is the inverse temperature and $H$ is the
Hamiltonian of the system. The partition function $Z$
gives all the statistical mechanics 
information
about the system, in particular its thermodynamics.
Through the Feynman path integral and by
performing the Euclideanization of time,
one obtains the path integral formula for the
partition function $Z = \int D\psi\, {\rm e}^{-I[\psi]}$, where $I$ is
the Euclidean action and the integration is done
over periodic fields
in the imaginary time variable in the case the fields $\psi$
are bosonic.

In the Euclidean path integral approach to quantum gravity, one
extends this formula to gravitational systems. The partition function
of curved space with matter
can then be given by $Z = \int Dg_{\alpha \beta}
D\psi\, {\rm e}^{-I[g_{\mu \nu}, \psi]}$, where $g_{\alpha \beta}$ 
represents the Euclidean metric, $\psi$ describes the matter
fields, and $I$ is the Euclidean action. The integration is done over
metrics and matter fields that are periodic in Euclidean
time, with fixed quantities at the boundary of the curved
space. When dealing with
the canonical ensemble, the data fixed at the
boundary are simply the inverse temperature $\beta = \frac{1}{T}$,
which is given by the total Euclidean proper time at the boundary. For
the case of asymptotically AdS spaces, the Euclidean proper
time length at infinity becomes zero and we instead use the asymptotic
behavior of AdS to fix a quantity
${\bar\beta}$, which is
proportional to the inverse of the temperature ${\bar T}$ of a conformal boundary.
We choose the proportionality constant to be one so that
${\bar\beta} = \frac{1}{{\bar T}}$.

Due to the difficulties in performing the full path integral, we
employ the zero-loop approximation of the path integral, but we do it
in a different way. We assume that we can put the path integral over
the matter fields inside the path integral over metrics in the sense
of $Z = \int Dg_{\alpha \beta}\, {\rm e}^{-I_{g}}
\int D\psi \,{\rm e}^{-I_\psi}$, where $I_g=I_{g}[g_{\mu \nu}]$
is the Euclidean gravitational action
and $I_\psi=I_\psi[g_{\mu \nu},\psi]$ is the
Euclidean matter action of any field $\psi$. We assume minimal
coupling between the field $\psi$ and the metric $g_{\alpha \beta}$.
While for the general case one cannot perform the path integral on
matter, for the case of a matter thin shell in spherical symmetry one
can perform the path integral exactly, if the action is quadratic in
the field. This is because the metric components will be constants in
the action of the matter thin shell and the path integral becomes an 
integration over Gaussian functions, yielding
$\int D\psi
\,{\rm e}^{-I_\psi[g_{\mu \nu},\psi]}=
{\rm e}^{-I_{\rm m}[g_{\mu \nu}]}$,
where $I_{\rm m}$ is an effective matter action. 
So, the partition function,
after performing the integration over the matter fields,
becomes
\begin{align}
Z = \int Dg_{\alpha \beta}\, {\rm e}^{-I[g_{\mu \nu}]}\,,
\label{eq:partitionfunctiongeneric}
\end{align}
with $I[g_{\mu \nu}]$ given by
\begin{align}
I=I_g + I_{\rm m}
\,,
\label{eq:actiongeneric}
\end{align}
more precisely, $I[g_{\mu \nu}]=I_g[g_{\mu \nu}] + I_{\rm
m}[g_{\mu \nu}]$.  This motivates the expression of the action
below. With the action, we perform the zero-loop approximation of the
path integral by first using the Hamiltonian and momentum constraints
to obtain the reduced action $I_*$. Then, the stationary points of the
reduced action are found and stability can be analyzed.

\subsection{Euclidean action}

The asymptotically AdS space can be represented by $M$, 
which splits into two spaces $M_1$ and $M_2$
due to the presence of a matter thin shell $N$ separating 
them. The outer boundary of $M$ lies at spatial infinity 
and is represented by $B$. The gravitational 
part of the Euclidean action considered here is the 
Euclidean Einstein-Hilbert action
with the Gibbons-Hawking-York boundary term written as
$I_g = - \frac{1}{16\pi l_{\rm p}^2}\int_{M\setminus\{N\}} 
   \left(R + \frac{6}{l^2}\right)\sqrt{g} d^4x 
   + \int_{N}  \frac{[K]}{8\pi l_{\rm p}^2} 
    \sqrt{h}d^3y 
   - \frac{1}{8\pi l_{\rm p}^2}\int_{B} K \sqrt{h}d^3y - I_{\rm subtr}
$,
where
$l_{\rm p}$ is the Planck length, 
$R$ is the Ricci scalar, $g$ is the metric determinant, 
$l = \sqrt{-\frac{3}{\Lambda}}$ is defined as the AdS length,
with $\Lambda$ 
being the negative
cosmological constant,
 $h_{ab}$ is the induced 
metric from the space metric $g_{\alpha \beta}$
on the hypersurface in analysis,
$h$ is the
determinant of $h_{ab}$,
$K_{ab}$ is the extrinsic curvature 
of the hypersurface in analysis, with 
trace $K$ given by $K= {n^\alpha}_{;\alpha}$,
$n^\alpha$ being the normal vector to the  hypersurface in analysis,
i.e., either $N$ or $B$,
the bracket $[K] = \eval{K}_{M_2}- \eval{K}_{M_1}$ 
means the difference  between 
$K$ evaluated at $M_2$ and $K$ evaluated at $M_1$,
and $I_{\rm subtr}$ is the action of a reference space, in 
this case AdS.
In relation to the matter part of the
Euclidean action,
we take as Lagrangian density,
the matter free energy per
unit area $\mathcal{F}_{\rm m}$.
This stems from the fact we
are dealing with the canonical ensemble,
which is connected to the thermodynamic Helmholtz free energy.
Then, 
$I_{\rm m}= \int \mathcal{F}_{\rm m}[h_{ab}] \sqrt{h}d^3x$,
where $\mathcal{F}_{\rm m}$
is a functional of the induced metric $h_{ab}$
on the shell, with $h$ being the 
determinant of $h_{ab}$. 
The Euclidean action of the system
$I=I_g+I_{\rm m}$ is then given 
by  
\begin{align}
   I = &- \frac{1}{16\pi l_{\rm p}^2}\int_{M\setminus\{N\}} 
   \left(R + \frac{6}{l^2}\right)\sqrt{g} d^4x 
   + \int_{N} \left( \frac{[K]}{8\pi l_{\rm p}^2} 
   + \mathcal{F}_{\rm m}[h_{ab}]\right)\sqrt{h}d^3y \notag\\
   & - \frac{1}{8\pi l_{\rm p}^2}\int_{B} K \sqrt{h}d^3y - I_{\rm subtr}\,,
   \label{eq:action}
\end{align}
with all quantities having been properly defined.

\subsection{Spherically symmetric geometry}

In the analysis
we perform of the path integral,
we only consider paths which are spherically
symmetric. We also assume that the space is static. We write then
the metric for $M_1$ as
\begin{align}
   & ds^2_{M_1} = 
   b_1^2(y) \frac{b^2_2(y_{\rm m})}{b_1^2(y_{\rm m})}d\tau^2 
   + a_1^2(y)dy^2 + r(y)^2 d\Omega^2\,,\quad\quad 0\leq y<y_{\rm m}\,.
\label{eq:metricM1}
\end{align}
For the thin shell $N$, we write the induced metric as
\begin{align}
   ds^2_{N} = b^2_2(y_{\rm m})d\tau^2 
   + r^2(y_{\rm m})d\Omega^2\,,\quad\quad
   \quad\quad\quad\quad\quad\quad y=y_{\rm m}\,.
   \label{eq:inducedmetricA}
\end{align}
For $M_2$, we write 
\begin{align}
   & ds^2_{M_2} = b^2_2(y) d\tau^2
    + a_2^2(y)dy^2 + r^2(y)d\Omega^2
   \,,\quad\quad\quad\quad\quad\quad y_{\rm m}< y<1\,.
\label{eq:metricM2}
\end{align}
Here $b_1$, $b_2$, $a_1$, $a_2$, and $r$ are functions of the
coordinate $y$. The
Euclidean time coordinate $\tau$ is chosen to be an angular
coordinate in the interval $0<\tau <2\pi$ on $M$, the
radial coordinate $y$
takes values as above, $d\Omega^2=d\theta^2+\sin^2\theta \,d\phi^2$ is
the line element of the $2$-sphere with surface area $\Omega = 4\pi$,
and the coordinates $\theta$ and $\phi$ are the usual spherical
coordinates.  The points at the thin shell are located at $y=y_{\rm
m}$ and we exhibit $y_{\rm m}$ as a label that is fixed, while the
radius of the shell $r(y_{\rm m})$ depends on the arbitrary function
$r(y)$.

\subsection{Matter free energy and stress-energy tensor}

The matter Lagrangian density can be identified to the thermodynamic 
Helmholtz free energy density since we are
dealing with the canonical ensemble.
The Helmholtz free energy potential $F$  is described by 
$F=E-TS$, where $E$ is the thermodynamic 
energy, $S$ the entropy, and $T$ the temperature of the reservoir.
We must then analyze the associated density quantities, and so the 
free energy density $\mathcal{F_{\rm m}}$ can be written as
\begin{align}
\mathcal{F_{\rm m}}[h_{ab}] = \epsilon_{\rm m}[h_{ab}] 
- T_{\rm m}[h_{ab}] s_{\rm m}[h_{ab}]
  \,,
   \label{eq:matterfe}
\end{align}
where $\epsilon_{\rm m}$ is the total energy density of the matter,
$T_{\rm m}$ is the local temperature of the shell, and 
$s_{\rm m}$ is the entropy density of the matter. All these quantities are 
functionals of the induced metric $h_{ab}$.
Since from Eq.~\eqref{eq:inducedmetricA},
one has that $h_{ab}$
depends on two arbitrary quantities that are seen as constants at the 
shell, 
$b_2(y_{\rm m})$ and $r(y_{\rm m})$,
the matter free energy density 
$\mathcal{F_{\rm m}}[h_{ab}]$ depends only locally
on these two quantities, and so a
dependence on derivatives of $h_{ab}$ is ruled out.

We define the radius $\alpha$ 
of the shell as
\begin{align}
   \alpha = r(y_{\rm m})\,.
   \label{eq:numberdensityshell1}
\end{align}
Although in order to keep a consistent nomenclature,
we should have defined the radius of the shell
as $r_{\rm m}=r(y_{\rm m})$,
we prefer to stick to
$\alpha = r(y_{\rm m})$ to not overcrowd the symbols ahead.
One can define a local temperature at some point $y$,
as $T(y) = \frac{1}{2\pi b_2(y)}$. So the local temperature of 
the shell is 
\begin{align}
   T_{\rm m} = \frac{1}{2\pi b_2(y_{\rm m})}\,.
   \label{eq:tempshell1}
\end{align}
The rationale for this definition comes from continuity, since in the
canonical ensemble one fixes the Euclidean proper time length at the
boundary and assigns to it the meaning of an inverse temperature. 
One must keep in mind however that this definition does not
give information about the specific expression of the temperature
since $b_2(y_{\rm m})$ is arbitrary.  

The free energy per unit area
$\mathcal{F}_{\rm m}[h_{ab}]=\mathcal{F}_{\rm m}[b_2(y_{\rm m}),
r(y_{\rm m})]$
can then be put in the form
\begin{align}
   &\mathcal{F}_{\rm m}[h_{ab}] = \mathcal{F}_{\rm{m}}
   [\alpha,T_{\rm m}]\,,
   \label{eq:freeenergyShell1}
\end{align}
upon using Eqs.~\eqref{eq:numberdensityshell1} and \eqref{eq:tempshell1}.
Now, we assume the first law to describe the matter energy  density 
as
$d \epsilon_{\rm m} = T_{\rm m} ds_{\rm m} -
2(\epsilon_{\rm m}
-T_{\rm m}s_{\rm m}
+ p_{\rm m})\frac{d\alpha}{\alpha}$,
where $p_{\rm m}$ 
is the matter tangential pressure at the shell.
Thus, from Eq.~\eqref{eq:matterfe}, the free energy density has
the differential,
\begin{align}
d \mathcal{F}_{\rm m} = - s_{\rm m} dT_{\rm m} -
2(\mathcal{F}_{\rm m} + p_{\rm m})
\frac{d\alpha}{\alpha}
\,.
\label{eq:difffreeenergy}
\end{align}
With the known differential of the free energy density regarding its
dependence on the metric components, one can compute the
surface stress-energy tensor $S^{ab}$
as the functional derivative, $S^{ab}
= - \frac{2}{\sqrt{h}} \frac{\delta (\sqrt{h}
\mathcal{F_{\rm m})}[h_{ab}]}{\delta h_{ab}}$.
From Eq.~\eqref{eq:inducedmetricA},
one has $h_{\tau\tau}=b^2_2(y_{\rm m})$
and $h_{\theta\theta}=\frac{h_{\phi\phi}}{\sin^2(\theta)}=r^2(y_{\rm
m})$.
Then, from
 $\alpha = r(y_{\rm m})$ and 
$T_{\rm m} = \frac{1}{2\pi b_2(y_{\rm m})}$,
see Eqs.~\eqref{eq:numberdensityshell1}
and \eqref{eq:tempshell1}, one finds that
the variation yields
 ${S^{\tau}}_{\tau} =
- \mathcal{F}_{\rm m} + T_{\rm m}
\frac{\p \mathcal{F}_{\rm m}}{\p T_{\rm m}}$
and 
 ${S^{\theta}}_{\theta} = {S^{\phi}}_{\phi} =
-\frac12\alpha\frac{\p \mathcal{F}_{\rm m}}{\p \alpha} -
\mathcal{F}_{\rm m}
$, where the partial derivatives are done keeping the
hidden variable constant,
and $\delta h= h h^{ab} \delta h_{ab}$ has been used. 
We note that some care is due while performing the variational
derivative to obtain ${S^{\theta}}_{\theta}$ and ${S^{\phi}}_{\phi}$,
as one
must calculate $d\alpha$ in $d\mathcal{F}_{\rm m}$ as 
$d\alpha = d\left(\sqrt[4]{\frac{h_{\theta\theta} 
h_{\phi \phi}}{\sin^2\theta}}\right)$.
From Eqs.~\eqref{eq:matterfe} and \eqref{eq:difffreeenergy},
one has  $\epsilon_{\rm m} = \mathcal{F}_{\rm m} - T_{\rm m}
\frac{\p \mathcal{F}_{\rm m}}{\p T_{\rm m}}$ and $p_{\rm m} =
-\frac12\alpha\frac{\p \mathcal{F}_{\rm m}}{\p \alpha} -
\mathcal{F}_{\rm m}$.
Thus, the stress-energy tensor ${S^a}_{b}$ has components,
\begin{align}
{S^{\tau}}_{\tau} = - \epsilon_{\rm m}\,,\quad\quad
{S^{\theta}}_{\theta} = {S^{\phi}}_{\phi} = p_{\rm m}\,.
\label{eq:stress-energytensor}
\end{align}
The fluid is thus isotropic, more specifically, it is
a perfect fluid. 
Note that $\epsilon_{\rm m}=\epsilon_{\rm m}(\alpha,T_{\rm m})$
and $p_{\rm m}=p_{\rm m}(\alpha,T_{\rm m})$.
The rest mass $m$ of the shell is important for the analysis below, and 
we define it as
\begin{align}
m= 4\pi \alpha^2
\epsilon_{\rm m}\,.
\label{eq:massshell}
\end{align}
Since $\epsilon_{\rm m}=\epsilon_{\rm m}(\alpha,T_{\rm m})$, one has
$m=m(\alpha,T_{\rm m})$. The dependence of the thermodynamic
quantities in $\alpha$ and $T_{\rm m}$ is helpful when one makes
variations of the action on the metric components to find the
Hamiltonian constraint. However, it is also helpful  to
invert the first law
of thermodynamics to get $ds_{\rm m} = \frac{1}{T_{\rm
m}}d\epsilon_{\rm m} + 2(\epsilon_{\rm m} - T_{\rm m} s_{\rm m} +
p_{\rm m}) \frac{d\alpha}{\alpha}$. We can integrate over the area
to obtain the first law of thermodynamics in the form, 
\begin{align}
T_{\rm m}dS_{\rm m} = dm+
p_{\rm m} dA_{\rm m},
\label{eq:1stlawintegral}
\end{align}
where
\begin{align}
A_{\rm m}= 4\pi \alpha^2,
\label{eq:areaAm}
\end{align}
\begin{align}
S_{\rm m}= s_{\rm m} A_{\rm m},
\label{eq:entropySm}
\end{align}
are the area of the shell and 
the entropy of the matter in the shell, respectively.
Written likes this, also
the quantities $S_{\rm m}$, $T_{\rm m}$, and $p_{\rm m}$ become
functions of $m$ and $\alpha$.  We shall explicitly indicate which
dependences are used below.

\subsection{Boundary conditions}

\subsubsection{Origin: $y=0$}

We must further impose regularity conditions and boundary conditions 
on the metrics that are being summed in the path integral. 
In the region $M_1$, the interior
region, we impose regularity conditions 
at $y=0$ corresponding to flat conditions at the origin, i.e.,
\begin{align}
   r|_{y=0}= 0\,,\quad\quad\
 b_1|_{y=0} \,\,{\rm finite\,\, and \,\,positive}\,,
\label{eq:regularitytopology}
\end{align}
\begin{align}
   \eval{\frac{r'}{a_1}}_{y=0} = 1\,,\,\quad
   \eval{\frac{1}{r'}\left(\frac{r'}{a_1}\right)'}_{y=0} = 0\,,
\quad
\eval{\frac{b_1'}{a_1}}_{y=0} = 0 \,,
   \label{eq:regularitycurvature}
\end{align}
where a primed quantity means derivative
with respect to $y$, e.g.,
$r' = \frac{dr}{dy}$. The
two conditions in Eq.~\eqref{eq:regularitytopology} 
are the result of choosing the topology $S^1 \times R^3$ for the region in 
$M_1$. The three conditions in 
Eq.~\eqref{eq:regularitycurvature} are needed so that the Ricci scalar is 
well-defined to avoid the occurrence of singularities in $M_1$.

\subsubsection{Infinity: $y=1$}

In the region $M_2$, the exterior region,
we impose that the space behaves as asymptotically 
AdS, when $y\rightarrow 1$.
The outer boundary $B$ is defined as the hypersurface with constant 
$y$ in the limit of $y\rightarrow 1$, 
with the radius $r(y)$ diverging in
this limit, so $y\to1$ implies
$r(y)\to\infty$.
Since AdS means a constant curvature space, 
the metric $g_{\alpha \beta}$ should 
satisfy asymptotically the vacuum Einstein equations
with negative cosmological 
constant, i.e.,
the Ricci tensor should obey $R_{\alpha \beta} = 
-\frac{3}{l^2}g_{\alpha \beta}$, and
the Ricci scalar should be $R=-\frac{12}{l^2}$.
To establish exactly the asymptotic behavior 
of the metric, one can perform a conformal transformation
such that $\bar{g}_{\alpha \beta} = W^2 g_{\alpha \beta}$, 
where $\bar{g}_{\alpha \beta}$ is the 
unphysical metric and $W$ is a conformal factor.
Following~\cite{Ashtekar:1984}, the conformal 
factor must vanish at the boundary of the conformal 
space but in such a way that the conformal metric is nonsingular. 
The only way to satisfy these conditions is if the 
conformal factor behaves as $W = c\, r^{-1}(y)$ near 
the boundary, where 
$c$ is a constant that we have chosen here as $c=1$.
If one makes a coordinate transformation $W = W(y)$,
the conformal metric near the hypersurface $B$, which is now
described by $W=0$, has in 
the neighborhood $\mathcal{N}(B)$
of $B$
the following corresponding
line element,
$
d\bar{s}^2|_{\mathcal{N}(B)} = 
   \frac{b_2^2(y)}{r^2(y)} d\tau^2 
   + \left(\frac{a_2(y)r(y)}{r'(y)}\right)^2 dW^2(y)
   + d\Omega^2$.
The asymptotic behavior of the metric is going to be chosen as 
boundary conditions to be imposed on the elements of 
$d\bar{s}^2$~\cite{Ashtekar:1984,Henneaux:1985}.
The fact that
the original metric $g_{\alpha \beta}$ should 
satisfy asymptotically the vacuum Einstein equations with
negative cosmological constant, 
$R_{\alpha \beta} = 
-\frac{3}{l^2}g_{\alpha \beta}$,
yields two conditions. 
One
condition 
is that the
conformal induced metric on ${\mathcal{N}}(B)$ can be written as 
$d\bar{s}^2 = d\bar{\tau}^2 + d\Omega^2$, where $\bar{\tau}$ is a 
particular Euclidean time coordinate that can be recast 
into the coordinate $\tau$ as $\bar{\tau} = \gamma\tau$
with $\gamma$ being a 
constant.
The other condition is
that $\bar{g}^{\alpha \beta}\nabla_{\alpha}W 
\nabla_{\beta}W = \frac{1}{l^2}$.
Imposing these two conditions to the conformal 
metric in $d\bar{s}^2$,
one gets two following boundary 
conditions that can be 
cast in terms of the original metric components.
The first boundary condition is
\begin{align}
 \eval{\frac{b_2(y)}{r(y)}}_{y\to1} = 
   \frac{{\bar\beta}}{2\pi l}\,,
   \label{eq:beta*}
\end{align}
where $\frac{{\bar\beta}}{2\pi l}$ 
is meant as the constant $\gamma$ 
of the first condition
above, and the parameter ${\bar\beta}$
is defined to be the fixed quantity of the ensemble. In some sense,
the parameter ${\bar\beta}$ is proportional to the total proper length of
the conformal boundary, and we identify it as being equal to the
inverse of the local temperature ${\bar T}$
of the conformal boundary,
such that 
${\bar\beta} = \frac{1}{{\bar T}}$.
It must be pointed out that fixing the inverse
temperature in this conformal boundary as ${\bar\beta}$ is a
choice of the formalism, with the only freedom being the 
choice of a constant of proportionality. Here, the choice coincides with the usual
Euclidean proper time approach formalism, e.g., the way we
define the temperature at infinity for a black hole yields the same as
the Hawking-Page definition. This can be seen further below.
The second boundary condition is
\begin{align}
\eval{\frac{a_2(y) r(y)}{r'(y)}}_{y\to1} = l
\,,
\label{eq:bounda2}
\end{align}
which gives the typical asymptotic behavior of $a_2$ in
asymptotically AdS space.

We note that one could have chosen a different constant of 
proportionality $c$ in the
conformal transformation as long as the asymptotic AdS behavior 
is imposed. This indeed leads to a 
${\bar\beta}$ only differing by a constant multiplication factor 
which can be thought of as a change of scale for the temperature 
and energy measured by the conformal observer at the boundary, 
e.g., a change of units in the measuring tools of the observer. 
The physical results do not alter from such choice.


\subsection{Hamiltonian and momentum constraint equations}

We impose the Hamiltonian and momentum constraint equations so that
the path integral is along the constraint paths.
In this context,  the 
Hamiltonian constraint arises from the first variation of the action 
with respect to the metric components $b_1(y)$ and $b_2(y)$, and 
so the application of the constraint to the paths can be
seen as a first step towards zero-loop approximation, establishing 
the relation between the curvature of space and the
energy density of matter, as in general relativity.
Moreover, this is
powerful enough that allows the analysis of stability. Only afterwards
we perform the full zero-loop approximation.  We start with the
Hamiltonian constraint, consisting of one equation for each region
$M_1$ and $M_2$, and a junction condition on the matter shell $N$.  We
then analyze the momentum constraint.

The Hamiltonian constraint 
in the regions $M_1$ and $M_2$ makes use of
the Einstein tensor component
${G^{\tau}}_{\tau}$, which can be written as
${G^{\tau}}_{\tau}|_{M_i} = 
   \frac{1}{r' r^2}
   \left(r\left[ \left(\frac{r'}{a_i}\right)^2 - 1\right]\right)'$
for each $M_i$, with $i=\{1,2\}$.
The Hamiltonian constraint is the 
Einstein equation ${G^{\tau}}_{\tau} = \frac{3}{l^2}$, or explicitly
$ \frac{1}{r' r^2}
   \left(r\left[ \left(\frac{r'}{a_i}\right)^2 - 1\right]\right)'
   = \frac{3}{l^2}$,
for each region $M_i$.  
Thus, the Hamiltonian constraint 
for each region is satisfied if 
\begin{align}
   & \left(\frac{r'}{a_1}\right)^2  = 
   1 + \frac{r^2}{l^2}\equiv f_1(r)\,,
   \label{eq:a1constraint}\\
   &  \left(\frac{r'}{a_2}\right)^2  = 
   1 + \frac{r^2}{l^2} 
   - \frac{\tilde{r}_+ + \frac{\tilde{r}_+^3}{l^2}}{r}
   \equiv f_2(r,\tilde{r}_+)\,,
   \label{eq:a2constraint}
\end{align}
where the regularity condition $\frac{r'}{a_1}\sVert[1]_{y=0} = 1$ in
Eq.~\eqref{eq:regularitycurvature} was used, $\tilde{r}_+$ is the
gravitational radius of the system and it is featured as an
integration constant obeying $\tilde{r}_+ < \alpha$, and we have
defined the functions $f_1(r)$ and $f_2(r,\tilde{r}_+)$ for
convenience. Due to the order of the differential equation in the
Hamiltonian constraint equation, the regularity condition
$\frac{1}{r'}\left(\frac{r'}{a_1}\right)'\sVert[2]_{y=0} = 0$ in
Eq.~\eqref{eq:regularitycurvature} was not used but it is naturally
satisfied.  The same thing happens for the function $\frac{r'}{a_2}$
which obeys naturally the boundary condition Eq.~\eqref{eq:bounda2}.

The Hamiltonian constraint in the hypersurface $N$ is described by the
junction condition $[{K^{\tau}}_{\tau}] - [K] = - 8\pi l_{\rm p}^2
{S^{\tau}}_{\tau}$, where ${S^{\tau}}_{\tau}$ is the $\tau\tau$
component of the surface stress-energy tensor. The extrinsic
curvature components of constant $y$ surfaces are given by
${K^{\tau}}_{\tau}|_{M_i} = \frac{b'_i}{a_i b_i}$, 
   and
  $ {K^{\theta}}_{\theta}|_{M_i} 
   = {K^{\phi}}_{\phi}|_{M_i} = \frac{r'}{a_i r} $, 
with the trace of the extrinsic curvature being 
$K = \frac{b'_i}{a_i b_i} + 2 \frac{r'}{a_i r}$.
The surface stress-energy tensor is the functional derivative
$S^{ab}$, with ${S^{\tau}}_{\tau}=-\epsilon_{\rm m}$, see
Eq.~\eqref{eq:stress-energytensor}.
Then, for the mass $m= 4\pi \alpha^2 \epsilon_{\rm m}$,
Eq.~\eqref{eq:massshell}, we find the Hamiltonian constraint at the
shell is
$m = \frac{\alpha}{l_{\rm p}^2} \eval{\left[\frac{r'}{a_1} 
   - \frac{r'}{a_2}\right]}_{y=y_{\rm m}}$, i.e.,
\begin{align}
   m = \frac{\alpha}{l_{\rm p}^2}(\sqrt{f_1(\alpha)} -
   \sqrt{f_2(\alpha,\tilde{r}_+)})
   \label{eq:junctioncondhamiltonian}
   \,,
\end{align}
with $f_1(\alpha)=\eval{\left(\frac{r'}{a_1}\right)^2}_\alpha
=1 + \frac{\alpha^2}{l^2}$ and
$f_2(\alpha,\tilde{r}_+)=
\eval{\left(\frac{r'}{a_2}\right)^2}_\alpha
=1 + \frac{\alpha^2}{l^2} 
   - \frac{\tilde{r}_+ + \frac{\tilde{r}_+^3}{l^2}}{\alpha}$,
   see
   Eqs.~\eqref{eq:a1constraint}
and \eqref{eq:a2constraint}, respectively.
While the dependence of $m$ in the metric components 
is described by $m = m(\alpha,T_{\rm m})$,
one can invert in order to $T_{\rm m}$
and get $T_{\rm m} = T_{\rm m}(m,\alpha)$. Using now the 
junction condition Eq.~\eqref{eq:junctioncondhamiltonian}, 
we obtain the temperature of the shell as a function of $\tilde{r}_+$ 
and $\alpha$ as $T_{\rm m} = T_{\rm
m}(m(\tilde{r_+},\alpha),\alpha)$, as long as the equation of
state $T_{\rm m}(m,\alpha)$ is provided. 

In relation to the momentum constraints, due to the spherical symmetry
of the metrics in Eqs.~\eqref{eq:metricM1} and~\eqref{eq:metricM2} and
the symmetry on translations in $\tau$, the momentum constraints 
in the regions $M_1$ and $M_2$
are satisfied a priori. Moreover, the momentum constraint at the shell is
satisfied since the matter shell stress tensor is diagonal as
$\mathcal{F}_{\rm m}$ is a functional only of $b_2$ and $\alpha$.

\subsection{Reduced action and the partition function from
the constrained path integral}

If one investigates the regularity conditions at the origin in
Eqs.~\eqref{eq:regularitytopology}
and \eqref{eq:regularitycurvature}, the boundary conditions
at infinity in
Eqs~\eqref{eq:beta*} and~\eqref{eq:bounda2}, the Hamiltonian
constraint solutions in
Eqs.~\eqref{eq:a1constraint},~\eqref{eq:a2constraint},
and~\eqref{eq:junctioncondhamiltonian}, and the fact that the Ricci
scalar here is given by $R|_{M_i} = -2
{G^{\tau}}_{\tau}\sVert[1]_{M_i} - \frac{2}{a_i b_i
r^2}\left(\frac{r^2 b'_i}{a_i}\right)'$ for each region $M_i$, the
action in Eq.~\eqref{eq:action} with these considerations yields the
reduced action $I_*$ with the form
$I_* = 
   -\frac{{\bar\beta}}{l_{\rm p}^2 l}
   \left(r^2 \sqrt{f_2(r)}\right)\sVert[2]_{y\rightarrow 1} - 
   S_{\rm m}(m(\tilde{r}_+,\alpha),\alpha) - I_{\rm subtr}$,
where an 
equation of state for the shell is 
assumed so that $S_{\rm m}(T_{\rm m}(m,\alpha), \alpha)
= S_{\rm m}(m,\alpha)$. 
The first term in $I_*$ is formally 
divergent, the reason being that we are in asymptotically AdS 
space. We must therefore pick the reference action $I_{\rm subtr}$ so that 
pure hot  AdS space
has vanishing action with the same 
inverse temperature ${\bar\beta}$ at the conformal boundary, thus 
regularizing the reduced action. 
Pure  hot  AdS space is defined 
as AdS at fixed ${\bar\beta}$ with nothing in it,
it 
has vanishing action, and could as well be called classical
hot AdS space. This notion 
of pure, or classical, hot AdS space, is very important
in the zero-loop approximation of the path integral
as it renders the approach self-consistent at zero-loop order.
Pure hot AdS space can be obtained by setting $\tilde{r}_+ = 0$ 
and $S_{\rm m} = 0$, and so $I_{\rm subtr}= 
-\frac{{\bar\beta}}{l_{\rm p}^2 l}\left(r^2
\sqrt{f_1(r)}\right)\sVert[2]_{y\rightarrow 1} $. 
By inspecting the limit $
\left(r^2 \sqrt{f_1(r)} - r^2\sqrt{f_2(r)}\right)\sVert[2]_{y\rightarrow 1} 
= \frac{l}{2}
(\tilde{r}_+ + \frac{\tilde{r}_+^3}{l^2})$,
the reduced action becomes
\begin{align}
   I_*[{\bar\beta};\tilde{r}_+,\alpha] 
   = \frac{{\bar\beta}}{2 l_{\rm p}^2}\left(\tilde{r}_+ +
   \frac{\tilde{r}_+^3}{l^2}\right) 
   - S_{\rm m}(m(\tilde{r}_+,\alpha),\alpha)
   \,,\label{eq:reducedaction1}
\end{align}
where $m(\tilde{r}_+,\alpha)$ is given by the right-hand side of
Eq.~\eqref{eq:junctioncondhamiltonian}. The partition function
of Eq.~\eqref{eq:partitionfunctiongeneric}
with its path integral reduces thus to the 
following expression, 
\begin{align}
   Z[{\bar\beta}] = \int D\tilde{r}_+ D\alpha 
   \,{\rm e}^{- I_*[{\bar\beta}; \tilde{r}_+,\alpha]}\,,
   \label{eq:reducedactionpathintegral}
\end{align}
as the sum over different metrics with spherical symmetry reduces to
the sum over metrics with different $\tilde{r}_+$ and different
$\alpha$.
For clarification, the Hamiltonian constraint that was 
imposed makes the dependence of the action in the metric components 
$b_1(y)$ and $b_2(y)$ disappear, and moreover
makes the components $a_1$ and $a_2$ 
be functional dependent on 
$\tilde{r}_+$ and on $r(y)$.
The integration over $\alpha$ arises due
to the sum over metric
functions $r(y)$. Although the Hamiltonian constraint
ensures that the metric in the bulk has the same form for any arbitrary 
function $r(y)$, through a coordinate transformation $r=r(y)$,
the value $\alpha = r(y_{\rm m})$ that
separates the regions $M_1$ and $M_2$ depends on the specific function
$r(y)$, and so one must sum over the possible values of $\alpha$.

\section{The zero-loop approximation and stability
criteria from the reduced action of a hot
self-gravitating thin shell in asymptotically AdS space
\label{sec:Zeroloopapproximation}}

\subsection{The stationary conditions, the
hot thin matter shell solutions in AdS space, and
the zero-loop approximation from the reduced action}

With the reduced action of the system being given by
Eq.~\eqref{eq:reducedaction1},
we now minimize it to find
the action in the zero-loop 
approximation. To find
the minimum of the action we have
to find its stationary conditions which
are given by
\begin{align}
\frac{\p I_*[{\bar\beta};\tilde{r}_+,\alpha]}{\p \alpha}= 0   \,,
\label{eq:stationcond1}
\end{align}
\begin{align}
\frac{\p I_*[{\bar\beta};\tilde{r}_+,\alpha]}{\p \tilde{r}_+}= 0\,.
\label{eq:stationcond2}
\end{align}
The stationary conditions given in Eqs.~\eqref{eq:stationcond1}
and~\eqref{eq:stationcond2} can be understood as the remaining
Einstein equations whose solutions minimize the action.  Since the
reduced action is essentially the Einstein-Hilbert action together
with the matter action, having the Hamiltonian constraint being
imposed, the minimization of the action in relation to $\alpha$ and to
$\tilde{r}_+$ is equivalent to the minimization in relation to the
metric components $g_{\theta \theta}$ and to $g_{yy}$. And so, these
conditions yield the Dirac delta terms of $G_{\theta \theta}=8 \pi
T_{\theta \theta}$, and the equation $G_{yy} = 0$, where $G_{\theta
\theta}$ and $G_{yy}$ are the corresponding components of the Einstein
tensor, and $T_{\theta \theta}$ is the corresponding component of the
stress-energy tensor.
In order to further develop the derivatives of
Eqs.~\eqref{eq:stationcond1} and \eqref{eq:stationcond2},
we have to find  $\frac{\p
S_{\rm m}}{\p \alpha}$ and $\frac{\p S_{\rm m}}{\p \tilde{r}_+}$.
We remind that, 
from the first law of thermodynamics
given in Eq.~\eqref{eq:1stlawintegral}, the matter entropy has
the differential form $dS_{\rm m} = 
\frac{dm(\tilde{r}_+,\alpha)}{T_{\rm m}} +
\frac{p_{\rm m}}{T_{\rm m}}dA_{\rm m}$,
and so to find the derivatives of $S_{\rm m}$
we have to find $\frac{\p A_{\rm m}}{\p \alpha}$,
$\frac{\p A_{\rm m}}{\p \tilde{r}_+}$,
$\frac{\p m}{\p \alpha}$, and
$\frac{\p m}{\p \tilde{r}_+}$.
From the expression of $A_{\rm m}$,
Eq.~\eqref{eq:areaAm}, we have
\begin{align}
\frac{\p A_{\rm m}}{\p \alpha} = 8\pi
\alpha, \quad\quad\quad \frac{\p A_{\rm m}}{\p \tilde{r}_+} = 0\,.
\label{eq:derivarea}
\end{align}
From the expression of $m(\tilde{r}_+,\alpha)$,
Eq.~\eqref{eq:junctioncondhamiltonian}, we have
\begin{align}
&\frac{\p m}{\p \alpha} \equiv - 8\pi \alpha p_{\rm g}\,,
\quad\quad\quad 
\frac{\p m}{\p \tilde{r}_+} = \frac{1 +
\frac{3\tilde{r}_+^2}{l^2}} {2 l_{\rm p}^2 \sqrt{f_2(\alpha, \tilde{r}_+)}}\,,
\nonumber
\\
&
p_{\rm g} \equiv \frac{1}{8\pi \alpha l_{\rm p}^2}
   \left(\frac{1 + 2 \frac{\alpha^2}{l^2}
   - \frac{\tilde{r}_+ + \frac{\tilde{r}_+^3}{l^2}}{2\alpha}}
   {\sqrt{f_2(\alpha, \tilde{r}_+)}} - 
   \frac{1 + 2\frac{\alpha^2}{l^2}}{\sqrt{f_1(\alpha)}}\right)
\,,
\label{eq:pg}
\end{align}
where $p_{\rm g}$ is defined as the gravitational pressure.
Using then the chain rule on the first law of thermodynamics,
we get the following 
derivatives for the matter entropy
\begin{align}
\frac{\p
   S_{\rm m}}{\p \alpha} = \frac{8\pi \alpha}{T_{\rm m}}(p_{\rm m} -
   p_{\rm g}), \quad\quad\quad 
    \frac{\p S_{\rm m}}{\p \tilde{r}_+} = \frac{1 + 3
   \frac{\tilde{r}_+^2}{l^2}} {2 l_{\rm p}^2 T_{\rm m}
   \sqrt{f_2(\alpha,\tilde{r}_+)}}\,.
   \label{eq:derTSm}
\end{align}
Since the differential of the entropy has been
recast in terms of  $\alpha$ and $\tilde{r}_+$, the temperature of the
shell and the pressure of the shell also have that dependence as
$T_{\rm m}= T_{\rm m}(m(\alpha,\tilde{r}_+),\alpha) =T_{\rm
m}(\alpha,\tilde{r}_+)$ and $p_{\rm m}=p_{\rm
m}(m(\tilde{r}_+,\alpha),\alpha) = p_{\rm m}(\alpha,\tilde{r}_+)$.
From now on, we abbreviate this dependence to avoid cluttering.
However, the dependence must always be assumed.

Then, using the reduced action given in Eq.~\eqref{eq:reducedaction1}
and the derivatives of the matter entropy in Eq.~\eqref{eq:derTSm}, one
finds the stationary conditions.
The stationary condition of 
Eq.~\eqref{eq:stationcond1} yields 
\begin{align}
p_{\rm g} = p_{\rm m}\,.
\label{eq:statio1}
\end{align}
This equation gives the condition for mechanical
equilibrium of the shell
and we call Eq.~\eqref{eq:statio1} as the balance of pressure
equation.
The stationary condition of 
Eq.~\eqref{eq:stationcond2}  yields 
\begin{align}
{\bar\beta} = \frac{1}{T_{\rm m} \sqrt{f_2(\alpha, \tilde{r}_+)}}\,.
   \label{eq:statio2}
\end{align}
This equation gives the condition for thermodynamic
equilibrium of the shell,
and we call Eq.~\eqref{eq:statio2}
as the balance of temperature equation.

One can
verify that Eq.~\eqref{eq:statio1} only depends on $\tilde{r}_+$ and 
$\alpha$, which means the solutions to this equation can 
be expressed as
\begin{align}
   \alpha=\alpha(\tilde{r}_+)\,.
   \label{eq:statio1final}
\end{align}
Then, one can input such solutions into
Eq.~\eqref{eq:statio2} to obtain an equation only dependent on
$\tilde{r}_+$ and ${\bar\beta}$, which can be solved by a function
$\tilde{r}_+({\bar\beta})$, i.e.,
\begin{align}
   {\bar\beta} = \frac{\iota(\tilde{r}_+)}
{1 + 3\frac{\tilde{r}_+^2}{l^2}}\,,\quad\quad
{\rm implying}
\quad\quad
\tilde{r}_+=\tilde{r}_+({\bar\beta})\,,
\label{eq:statio2final}
\end{align} 
where $\iota(\tilde{r}_+)$
is a function of $\tilde{r}_+$ that appears for
convenience and is given for the thin shell by
\begin{align}
\iota(\tilde{r}_+)=\frac{1 + 3\frac{\tilde{r}_+^2}{l^2}}
   {T_{\rm m}(\tilde{r}_+,\alpha(\tilde{r}_+))
   \sqrt{f_2(\alpha(\tilde{r}_+))}}.
\label{eq:iota}
\end{align}
In comparison, for the Hawking-Page black hole one has
$\iota(\tilde{r}_+)=4\pi r_+$.
Of course, the expression
of Eq.~\eqref{eq:iota} means we are dealing with a thin shell.

Since, from Eq.~\eqref{eq:statio1final}, one has
$\alpha=\alpha(\tilde{r}_+)$,
the reduced action $I_*[{\bar\beta};\alpha,\tilde{r}_+]$
of Eq.~\eqref{eq:reducedaction1} under the mechanical stationarity 
condition
can be written as an effective reduced action
of the form
$I_*[{\bar\beta};\tilde{r}_+]$.
This in turn implies that
the partition function 
given in Eq.~\eqref{eq:reducedactionpathintegral}
is now 
 $Z[{\bar\beta}] = \int D\tilde{r}_+
\,{\rm e}^{- I_*[{\bar\beta}; \tilde{r}_+]}$, 
as the zero-loop approximation 
in the path integral over $\alpha$ as been done, i.e.,
using the reduced
action evaluated at the  stationary point
provided by Eq.~\eqref{eq:statio1}, or what amounts
to the same thing, by Eq.~\eqref{eq:statio1final}.
It is interesting to note that this behavior can be deduced from the
structure of the reduced action in Eq.~\eqref{eq:reducedaction1}
together with Eq.~\eqref{eq:reducedactionpathintegral}. In fact, the 
path integral over $\alpha$ in the partition function, 
$\int D\alpha \,{\rm e}^{S_{\rm m}}$, corresponds to
the partition function of the 
microcanonical ensemble.
Therefore, this
indicates that the canonical ensemble of the full system can be
described by an effective reduced action determined by the
microcanonical ensemble of a hot
self-gravitating matter thin shell with
fixed $\tilde{r}_+$, $I_*[{\bar\beta};\tilde{r}_+,\alpha(\tilde{r}_+)]=
I_*[{\bar\beta};\tilde{r}_+]$, while the solutions $\alpha(\tilde{r}_+)$ are 
but a consequence of performing the zero-loop approximation 
on the path integral over $\alpha$, i.e., of performing the zero
loop approximation
on the microcanonical ensemble.
Given Eqs.~\eqref{eq:reducedaction1}, \eqref{eq:statio1final},
and \eqref{eq:statio2final}, we see that 
the solutions $\tilde{r}_+({\bar\beta})$ of the
canonical ensemble are the solutions of the problem.
Having $\tilde{r}_+({\bar\beta})$, one finds
$\alpha=\alpha(\tilde{r}_+({\bar\beta}))$,
and then the action $I_*[{\bar\beta};\tilde{r}_+({\bar\beta})]$,
which is the action of the stationary points.
This action is the zero-loop approximation action $I_0({\bar\beta})$.
Indeed, 
\begin{align}
I_{0}[{\bar\beta}] \equiv
I_*[{\bar\beta};\tilde{r}_+({\bar\beta})],
\label{eq:expandaction}
\end{align}
i.e., the zero-loop action
$I_{0}[{\bar\beta}]$ is found by evaluating the reduced
action around its stationary points with $\alpha(\tilde{r}_+({\bar\beta}))$
and $\tilde{r}_+({\bar\beta})$ being found from 
the stationary conditions.
The partition function of the canonical ensemble can then be obtained to be
\begin{align}
    Z({\bar\beta})={\rm e}^{- I_{0}[{\bar\beta}]} \,,
   \label{eq:zerolooppartitionfunction}
\end{align}
in the zero
loop approximation, and the thermodynamic properties of the system can 
be extracted.

\subsection{The stability criteria from the reduced action of a
hot self-gravitating thin shell in asymptotically AdS space}

Going a step further within this formalism,
we can apply the zero-loop approximation of the path integral in
Eq.~\eqref{eq:reducedactionpathintegral},
and go one order up to first order approximation
by evaluating the reduced
action around its stationary points up until second order and
write
\begin{align}
   I_*[{\bar\beta}; \tilde{r}_+,\alpha] = I_{0}[{\bar\beta}]
   + \sum_{ij} H_{ij} \delta r^i \delta r^j\,,
   \label{eq:expandaction2}
\end{align}
where $I_{0}[{\bar\beta}] = 
I_*[{\bar\beta};\tilde{r}_+({\bar\beta}),\alpha({\bar\beta})]$
is the reduced action evaluated at the stationary points
given in Eq.~\eqref{eq:expandaction},
with 
$\tilde{r}_+({\bar\beta})$ and $\alpha({\bar\beta})$
being found from 
the stationary conditions of $I_*$,
and $H_{ij} = \eval{\frac{\p^2 I_*}
{\p r^i \p r^j}}_0$ is the Hessian of the
reduced action $I_*$ evaluated at the 
stationary points, 
with the parameters $r^i = (\alpha,\tilde{r}_+)$, with $i = 1,2$. 
The partition function can then 
be written in the saddle point approximation as
\begin{align}
   Z[{\bar\beta}] = {\rm e}^{- I_{0}[{\bar\beta}]} \int D r^i 
   {\rm e}^{- \sum_{jk} H_{jk} \delta r^j \delta r^k}
   \label{eq:Z1}\,,
\end{align} 
where the first and second factors are the zero and first-loop
contributions. 
Although we only consider the zero-loop contribution,
i.e., the zero-loop approximation,
we also take into account the first-loop contribution in the sense that 
it gives us some information about the stability of the approximation. For the
path integral to converge and so for the formalism to be stable, the
Hessian $H_{ij}$ must be positive definite, i.e.,
the stationary points must correspond to a local minimum
of the reduced action.

The components of the Hessian are
\begin{align}
   & H_{\alpha\alpha} = \frac{8\pi \alpha}{T_{\rm m}}
   \left(\left(\frac{\p p_{\rm g}}{\p \alpha}\right)_{\tilde{r}_+} 
   - \left(\frac{\p p_{\rm m}}{\p \alpha}\right)_{\rm m} 
   + 8 \pi \alpha p_{\rm m}
\left(\frac{\p p_{\rm m}}{\p m} \right)_\alpha \right)\,
   \,,\label{eq:I*alphaalpha}
\end{align}
\begin{align}
   & H_{\alpha\,\tilde{r}_+} = 
   \left(\frac{1 +  \frac{3\tilde{r}_+^2}{l^2}}
   {2 T_{\rm m}\sqrt{f_2} l_{\rm p}^2}\right)
   \left(
   \frac{\frac{\alpha}{l^2} + \frac{\tilde{r}_+
   +
   \frac{\tilde{r}_+^3}{l^2}}{2\alpha^2}}
   {f_2} - 8\pi \alpha
   \left(\frac{\p p_{\rm m}}{\p m}\right)_{\alpha}\right)
   \,,\label{eq:I*rpalpha}
\end{align}
\begin{align}
    H_{\tilde{r}_+\tilde{r}_+} = 
   \left(\frac{1 + \frac{3\tilde{r}_+^2}{l^2}}
   {2 T_{\rm m} \sqrt{f_2} l_{\rm p}}\right)^2
   \left[\frac{1}{l_{\rm p}^2}\left(\frac{\p T_{\rm m}}{\p m}\right)_\alpha 
   - \frac{T_{\rm m}}{\alpha \sqrt{f_2}}\right]\,,
   \label{eq:I*rprp}
\end{align}
where
\begin{align}\label{eq:derpressure}
   \left(\frac{\p p_{\rm g}}{\p \alpha}\right)_{\tilde{r}_+} = 
   \frac{1}{8\pi \alpha^2 l_{\rm p}^2}
   \left(\frac{3\frac{\tilde{r}_+ +
   \frac{\tilde{r}_+^3}{l^2}}{2\alpha}
   \left(1 - \frac{\alpha^2}{l^2}\right) - 1 
   - 3\frac{\left(\tilde{r}_+ +
   \frac{\tilde{r}_+^3}{l^2}\right)^2}{4\alpha^2}}{f_2^{3/2}}
   + \frac{1}{f_1^{3/2}}\right)\,\,,
\end{align}
and since here we are
working with three variables, we have explicitly put
the variable that is kept constant in
the subscript of the parenthesis of the partial derivative. The sufficient
conditions for the positive definiteness of the Hessian are chosen to
be
\begin{align}
    H_{\alpha \alpha} > 0 \,,
    \label{eq:stab1}
\end{align}
\begin{align}
    H_{\tilde{r}_+ \tilde{r}_+} 
   - \frac{H_{ \alpha\,\tilde{r}_+}^2}{H_{\alpha \alpha}} > 0\,.
\label{eq:stab2}
\end{align}
Applying Eqs.~\eqref{eq:I*alphaalpha}-\eqref{eq:I*rprp}
to Eqs.~\eqref{eq:stab1} and \eqref{eq:stab2}, and including
the marginal case, one has
\begin{align}
 \left(\frac{\p p_{\rm g}}{\p \alpha}\right)_{\tilde{r}_+} 
   - \left(\frac{\p p_{\rm m}}{\p \alpha}\right)_{\rm m} 
   + 8 \pi \alpha p_{\rm m}
\left(\frac{\p p_{\rm m}}{\p m} \right)_\alpha \geq0\,,
    \label{eq:stab1final}
\end{align}
\begin{align}
\frac{d
\tilde{r}_+}{d {\bar T}}\geq0\,,
    \label{eq:stab2final}
\end{align}
respectively.
Indeed,  one can obtain the derivative of the 
solution $\tilde{r}_+({\bar\beta})$ 
by applying the derivative of ${\bar\beta}$ to 
Eqs.~\eqref{eq:statio1} and~\eqref{eq:statio2}, obtaining 
$\frac{d \tilde{r}_+}{d {\bar\beta}} = - \frac{1}{2 l_{\rm p}^2}\left( 1 + 3
\frac{\tilde{r}_+^2}{l^2}\right)
\left(H_{\tilde{r}_+ \tilde{r}_+}
- \frac{H^2_{\tilde{r}_+ \alpha}}
{H_{\alpha \alpha}} \right)^{\hskip-0.15cm -1}$.
And so Eq.~\eqref{eq:stab2} implies that $
\frac{d
\tilde{r}_+}{d {\bar\beta}}<0$, or in terms of temperature
$\frac{d
\tilde{r}_+}{d {\bar T}}>0$, leading to Eq.~\eqref{eq:stab2final},
when one includes the marginal case.
Regarding the meaning of these stability conditions, one can verify
that Eq.~\eqref{eq:stab1final} is precisely the mechanical stability
condition for a static shell in AdS with constant $\tilde{r}_+$.
Regarding the other condition in Eq.~\eqref{eq:stab2final}, in some
sense it is a thermal stability condition.

Another comment that we can make
is in regard to $\frac{d \alpha}{d {\bar\beta}}$.
One can also obtain the derivative of the 
solution $\alpha({\bar\beta})$ 
by applying the derivative of ${\bar\beta}$ to 
Eqs.~\eqref{eq:statio1} and~\eqref{eq:statio2}, obtaining 
$\frac{d \alpha}{d {\bar\beta}} = -
\frac{H_{\tilde{r}_+ \alpha}}{H_{\alpha \alpha}} \frac{d\tilde{r}_+}{d{\bar\beta}}$,
i.e., 
$\frac{d \alpha}{d {\bar T}} = -
\frac{H_{\tilde{r}_+ \alpha}}{H_{\alpha \alpha}} \frac{d\tilde{r}_+}{d{\bar T}}$.
Thus, if mechanical stability holds, $H_{\alpha \alpha} > 0$,
Eq.~\eqref{eq:stab1},
then the radius of
the shell $\alpha$ decreases with ensemble
temperature if $H_{\tilde{r}_+ \alpha}>0$, and
increases if $H_{\tilde{r}_+ \alpha}<0$. The sign
of $H_{\tilde{r}_+ \alpha}$ depends on the particular
shell one is studying.

\section{Thermodynamics of the hot self-gravitating thin shell 
in the zero-loop approximation\label{sec:thermodynamics}}

In the
statistical mechanics formalism of the 
canonical ensemble, the partition 
function is given by the free energy $F$ as 
$Z= {\rm e}^{-{\bar\beta} F}$, while the zero-loop approximation 
gives a partition function $Z= {\rm e}^{-I_0}$.
By connecting both, we have
$F = {\bar T} I_0$, where ${\bar T}$ is the temperature
of the system,  ${\bar T}=\frac{1}{{\bar\beta}}$.
Then, from
Eq.~\eqref{eq:reducedaction1}, the free energy is
\begin{equation}
F =\frac{\tilde{r}_+ \left( l^2 + \tilde{r}_+^2
\right)}{2l^2 l_{\rm p}^2} - {\bar T} S_{\rm m}\,,
\label{eq:freeenergy}
\end{equation}
with $\tilde{r}_+$
given by
the solution
$\tilde{r}_+=\tilde{r}_+({\bar T})$
of Eq.~\eqref{eq:statio2final},
and $\alpha$ given by
the solution
$\alpha=\alpha(\tilde{r}_+({\bar T}))$ of Eq.~\eqref{eq:statio1final}
together
with Eq.~\eqref{eq:statio2final}.

One can now obtain the thermodynamic quantities
for the system from the derivatives 
of the free energy. In terms of the thermodynamic
energy $E$, the temperature ${\bar T}$, and the entropy $S_{\rm m}$, the
free energy of a system 
and its 
differential are given by 
\begin{align}
F= E - {\bar T} S\,,\quad\quad\quad dF = - S d{\bar T}\,,
\label{eq:F1}
\end{align}
respectively.
From the Eqs.~\eqref{eq:freeenergy} and \eqref{eq:F1},
we obtain 
that the entropy of the system is 
\begin{align}
  S = S_{\rm m}\,,
  \label{eq:entropy}
\end{align}
and the mean energy is
\begin{equation} 
E = \frac{1}{2 l_{\rm p}^2}
\tilde{r}_+ \left( 1 + \frac{\tilde{r}_+^2}{l^2} \right)\,,
\label{eq:energy}
\end{equation}
with $\tilde{r}_+=\tilde{r}_+({\bar T})$.
One can identify the Schwarzschild-AdS mass $M$ 
as the right-hand side of Eq.~\eqref{eq:energy}, i.e.,  
$E = M$. 

Regarding thermodynamic stability,  
one must verify if the heat capacity
$C$
is positive. If 
\begin{equation}
C \geq0,
\label{eq:heatcapacity>0}
\end{equation}
the system is thermodynamically stable,
where we include the limiting case, otherwise it
is unstable.
The heat capacity is defined as
$C =  \frac{d E}{d {\bar T}}$.
Using Eq.~\eqref{eq:energy} together with 
Eq.~\eqref{eq:statio2final},
we find 
\begin{equation}
C = \frac{1}{l_{\rm p}^2}\frac{ \left( 1 +
\frac{3\tilde{r}_+^2}{l^2} \right) \iota^2(\tilde{r}_+) }{
\frac{12\tilde{r}_+
\iota(\tilde{r}_+)}{l^2} - 2 \left( 1 +
\frac{3\tilde{r}_+^2}{l^2} \right) 
\frac{\partial \iota(\tilde{r}_+)}{\partial \tilde{r}_+} }\,,
\label{eq:heatcapacity}
\end{equation}
where $\iota(\tilde{r}_+)=\frac{1 + 3\frac{\tilde{r}_+^2}{l^2}}
{T_{\rm m}(\tilde{r}_+,\alpha(\tilde{r}_+))
\sqrt{f_2(\alpha(\tilde{r}_+))}}$, see 
Eq.~\eqref{eq:iota}, and $\tilde{r}_+=\tilde{r}_+({\bar T})$.
We see that the heat capacity is positive if
\begin{align}
6\frac{\tilde{r}_+}{l^2} - \iota'(\tilde{r}_+) {\bar T}\geq 0\,,
\label{eq:stabheta} 
\end{align}
where we have included the limiting case.
Using Eq.~\eqref{eq:statio2final}, one finds that Eq.~\eqref{eq:stabheta}
is equivalent to $\frac{d\tilde{r}_+}{d {\bar T}}\geq0$ which is
Eq.~\eqref{eq:stab2final} of the ensemble theory.  We now see that the
remaining stability condition Eq.~\eqref{eq:stab1} is not present or
cannot be accessed by the thermodynamics of the system.
It is moreover interesting to see that the
thermodynamic stability of the canonical ensemble given by
Eqs.~\eqref{eq:heatcapacity>0}-\eqref{eq:stabheta} is
also given by the saddle point
approximation of the effective reduced
action  $I_*[{\bar\beta};\tilde{r}_+]$, only dependent in the parameter
$\tilde{r}_+$. We suppose that this is due to the fact that
$\tilde{r}_+$ is associated to the quasilocal energy and so the
effective reduced
action $I_*[{\bar\beta};\tilde{r}_+]$
plays the role of the appropriate generalized free
energy that when minimized yields indeed the thermodynamic equilibrium
and stability of the canonical ensemble. It is important to note also
that the thermodynamic stability of the canonical ensemble is
different from the intrinsic stability of the system in the sense of
Callen, as intrinsic stability requires more conditions on the
concavity of the free energy.

\section{Specific case of matter thin shell with 
barotropic equation of state
\label{sec:unspecifiedeos}}

In order to proceed with the analysis of the
canonical ensemble of a self-gravitating 
matter thin shell, we must now
give the equations of state for the matter
in the shell. Here we give an
equation of state for the pressure
in the form of a barotropic equation, i.e.,
\begin{align}
  p_{\rm m}(m,\alpha) = \frac13\frac{m}{4\pi\alpha^2}\,.
  \label{eq:eospressure}
\end{align}
The equation of state for
the temperature of the matter is chosen as
\begin{align}
  T_{\rm m} = \frac{4}{3 C_0} \frac{m^{\frac{1}{4}}}
  {\left( 4\pi \alpha^2\right)^{\frac{1}{4}}}\,,
   \label{eq:tempspecific}
\end{align}
where $C_0$ is a constant with units.
Then, integrating the first law of thermodynamics yields
that the matter
entropy has the equation 
\begin{align}
  S_{\rm m} = C_0 m^{\frac{3}{4}} (4\pi \alpha^2)^{\frac{1}{4}}\,.
  \label{eq:entropyspecific}
\end{align}
A more general equation for $p_{\rm m}(m,\alpha)$ in
Eq.~\eqref{eq:eospressure}
could be chosen,
e.g., $p_{\rm m}(m,\alpha) = \lambda\frac{m}{4\pi\alpha^2}$,
where $\lambda$ is a constant,
with $\lambda = \frac{1}{2}$ corresponding to the barotropic 
equation of state of a two-dimensional ultrarelativistic 
gas.
The
mechanical stability condition, Eq.~\eqref{eq:stab1final},
requires that  $\lambda< \frac{1}{2}$.
Since the dominant energy condition requires 
$-1<\lambda<1$, a reasonable range for $\lambda$ is 
$0 <\lambda< \frac{1}{2}$, as we require the pressure to be 
positive.
If we integrate this general expression
for the pressure, using from the first law of thermodynamics
that $p_{\rm m} = - 
\left(\frac{\p m}{\p A_{\rm m}}\right)_{S_{\rm m}}$, the
partial derivative in relation the
area $A_{\rm m}$ being defined at constant matter entropy,
we would obtain the expression 
for $S_{\rm m}$ in the form
$S_{\rm m}(m,\alpha) = S_{\rm m}(m(4\pi \alpha^2)^{\lambda})$, i.e.,
$S_{\rm m}$ is an arbitrary function of $m(4\pi \alpha^2)^{\lambda}$.
We could pick a power-law expression for the entropy
of the form 
$S_{\rm m}(m,\alpha) = C_0 m^{\delta} (4\pi \alpha^2)^{\delta \lambda}$,
with $C_0$ a constant and $\delta$ a number.
Then, the temperature
equation of state
could be deduced using the first law of thermodynamics,
i.e., 
$\frac1{T_{\rm m}}= 
\left(\frac{\p S_{\rm m}}{\p m}\right)_{A_{\rm m}}$
giving $T_{\rm m} =
\frac{m^{1-\delta}}{\delta C_0 (4\pi \alpha^2)^{\delta \lambda}}$.
We could instead
have picked up this equation of state for 
$T_{\rm m}$ from the necessity of the equality of the second order
cross derivatives to have an exact $S_{\rm m}$, and then
integrate the first law to find $S_{\rm m}$ itself.
We have narrowed
the analysis to
specific $\lambda$ and $\delta$, using 
Eqs.~\eqref{eq:eospressure}-\eqref{eq:entropyspecific}.
Namely, we have adopted $\lambda = \frac{1}{3}$
as given in Eq.~\eqref{eq:eospressure},
and which is in the interval $0 <\lambda< \frac{1}{2}$.
We have also singled out
$\delta = \frac{3}{4}$, so that the
temperature and the entropy have
equations given in 
Eq.~\eqref{eq:tempspecific} and
Eq.~\eqref{eq:entropyspecific}, respectively.
For numerical purposes, it is best to write the pressure as 
$l_{\rm p}^2 l
p_{\rm m}(m,\alpha) =
\frac13 
\frac{m\frac{l_{\rm p}^2}{l}}{4\pi\frac{\alpha^2}{l^2}}
$, the temperature as
$T_{\rm m} = \frac1l \frac{4}{3c_0}
\left(\frac{l}{l_{\rm c}}\right)^\frac14
\left(\frac{l}{l_{\rm p}}\right)^\frac12
\left(\frac{m l_{\rm p}^2}{l}\right)^{\frac{1}{4}} \left( 4\pi
\frac{\alpha^2}{l^2}\right)^{-\frac{1}{4}}$, and
the 
entropy as $S_{\rm m} = c_0 \left(\frac{l_{\rm c}}{l}\right)^\frac14
\left(\frac{l}{l_{\rm p}}\right)^\frac32
\left(\frac{m \,l_{\rm p}^2}{l}\right)^{\frac{3}{4}}
\left(4\pi \frac{\alpha^2}{l^2}\right)^{\frac{1}{4}}$, i.e., $C_0 =
c_0 l_{\rm c}^{\frac{1}{4}}$, where $c_0$ has no units
and $l_{\rm c}$ can be
understood as the Compton wavelength associated to the rest mass of
the constituents of the shell.
The motivation for the matter equations of state given above is
both physical and mathematical.  Physically, the equations of state
resemble the equations of state of a radiation gas. Namely, the
equation of state for the pressure $p_{\rm m}(m,\alpha)$ can be
thought of as the equation of state of a three-dimensional radiation
gas confined in a very thin shell of small width $l_{\rm c}$. It can also be
thought of as a two-dimensional gas of a fundamental field with some
Compton wavelength $l_{\rm c}$, which is the view we take here.
Mathematically, it allows for an analytical
treatment of the balance of the pressure which facilitates the search
for the solutions of the shell and the analysis of its stability.

\begin{figure}[h]
   \centering
   \includegraphics[width=0.60\textwidth]{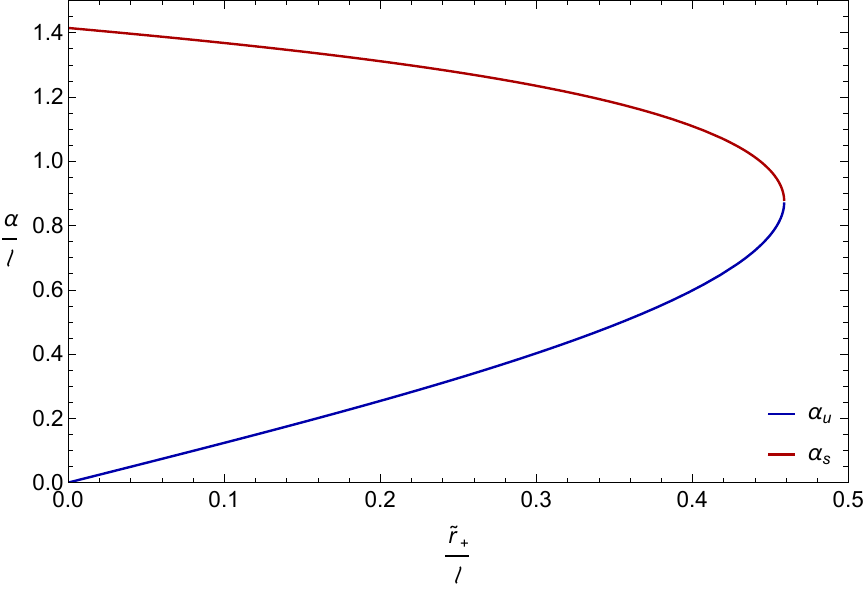}
   \caption{\label{fig:alphainrp}Solutions of the balance of pressure 
   $\frac{\alpha_{\rm u}}{l}$ and $\frac{\alpha_{\rm s}}{l}$
   as function of $\frac{\tilde{r}_+}{l}$.}
\end{figure}

With the equations of state described, we can solve numerically 
Eq.~\eqref{eq:statio2}, to obtain two solutions for the
radius $\alpha$ of the shell which we
write as $\alpha_{\rm u}(\tilde{r}_+)$
and $\alpha_{\rm s}(\tilde{r}_+)$,
see Fig.~\ref{fig:alphainrp}, the meaning of the subscripts u and s
will turn up shortly.
We see that the solution
$\alpha_{\rm u}(\tilde{r}_+)$ 
is monotonically increasing, 
while $\alpha_{\rm s}(\tilde{r}_+)$ 
is monotonically decreasing, until both meet 
a common point. By evaluating the stability condition 
in Eq.~\eqref{eq:stab1final},
it turns out that the solution
$\alpha_{\rm u}(\tilde{r}_+)$  is 
mechanically unstable, while
$\alpha_{\rm s}(\tilde{r}_+)$  is mechanically stable, hence the 
nature of the subscripts for the solutions.
This mechanical
behavior of the thin shell
is rather like the radius-mass behavior of a white dwarf
or a neutron star.
The two solutions translate into two possible radii
for the shell for a given energy, one which is large and another which
is small.  The small radius solution has very high pressure and is
unstable, the large radius solution has low pressure and is stable.
Physically, this is similar to the two solutions appearing in models
of astrophysical objects, such as white dwarfs, neutron stars
with polytropic-type equations of state, one solution being unstable
while the other being stable.

\begin{figure}[h]
   \centering
   \includegraphics[width=0.60\textwidth]{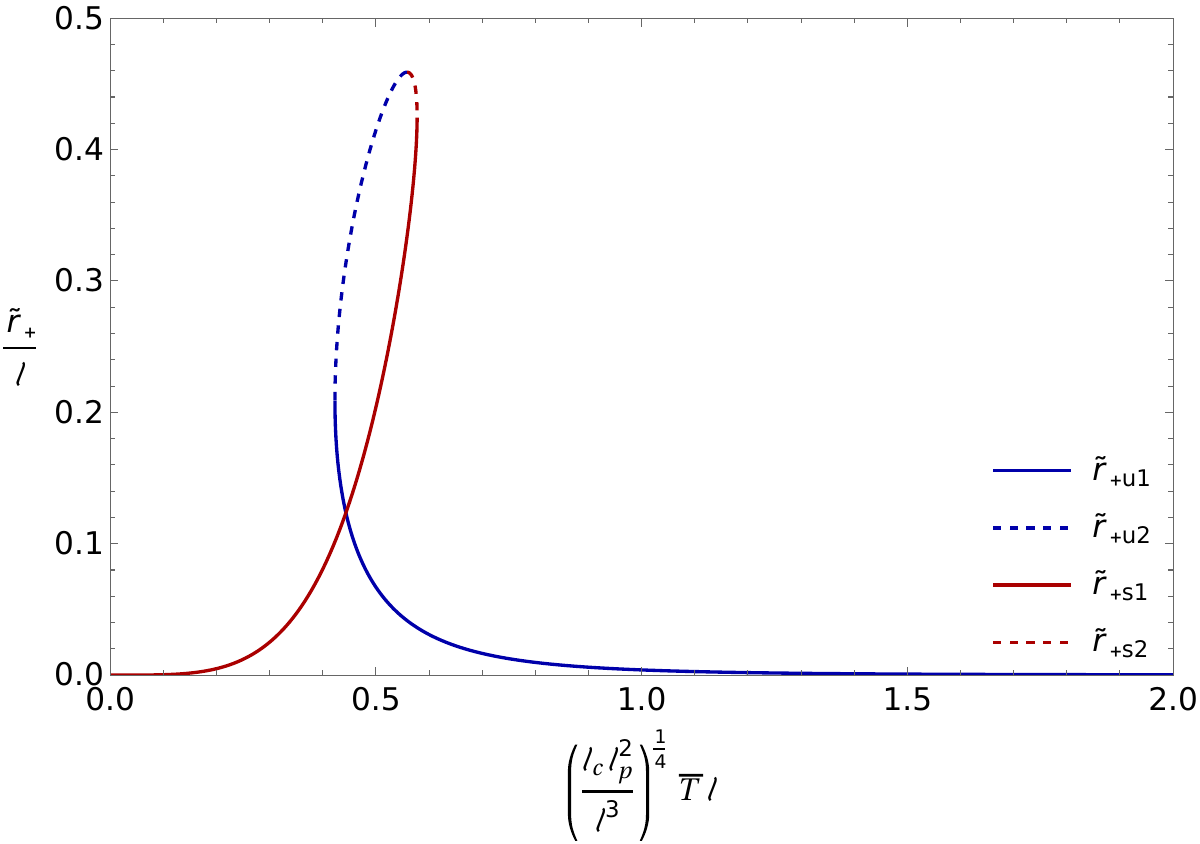}
   \caption{\label{fig:rpintstar} Solutions of the ensemble 
$\frac{\tilde{r}_{+{\rm u}1}}{l}$,
$\frac{\tilde{r}_{+{\rm u}2}}{l}$,
$\frac{\tilde{r}_{+{\rm s}1}}{l}$, and 
$\frac{\tilde{r}_{+{\rm s}2}}{l}$,
as functions of
${\bar T}l
\left(\frac{l_{\rm c}}{l}\right)^\frac14
\left(\frac{l_{\rm p}}{l}\right)^\frac12$.
Both
$\frac{\tilde{r}_{+{\rm u}1}}{l}$ and
$\frac{\tilde{r}_{+{\rm u}2}}{l}$
have shell radius $\alpha_{\rm u}$,
while both
$\frac{\tilde{r}_{+{\rm s}1}}{l}$ and
$\frac{\tilde{r}_{+{\rm s}2}}{l}$
have shell radius $\alpha_{\rm s}$.}
\end{figure}

Knowing the solutions
$\alpha_{\rm u}(\tilde{r}_+)$ and
$\alpha_{\rm s}(\tilde{r}_+)$
for the radius of the shell, we can put them
into Eq.~\eqref{eq:statio2} in order to obtain the solutions
for the gravitational
radius $\tilde{r}_+({\bar T})$.  One finds that there are
four solutions in total,
two solutions
$\tilde{r}_{+{\rm u}1}({\bar T})$ and
$\tilde{r}_{+{\rm u}2}({\bar T})$  with
shell radius
$\alpha_{\rm u}$, i.e.,
$\alpha_{\rm u} (\tilde{r}_{+{\rm u}1}({\bar T}))$ and
$\alpha_{\rm u} (\tilde{r}_{+{\rm u}2}({\bar T}))$, respectively,
and other two solutions
$\tilde{r}_{+{\rm s}1}({\bar T})$ and
$\tilde{r}_{+{\rm s}2}({\bar T})$  with shell radius
$\alpha_{\rm s}$, i.e., with
$\alpha_{\rm s}(\tilde{r}_{+{\rm s}1}({\bar T}))$
and
$\alpha_{\rm s}(\tilde{r}_{+{\rm s}2}({\bar T}))$,
respectively.
These solutions are
shown in Fig.~\ref{fig:rpintstar}. Regarding stability, the solutions
are thermodynamically stable if the gravitational radius increases
with the temperature ${\bar T}$.
From the figure,
we see that
$\tilde{r}_{+{\rm u}1}({\bar T})$ 
and $\tilde{r}_{+{\rm s}2}({\bar T})$
are thermodynamically unstable, and 
$\tilde{r}_{+{\rm u}2}({\bar T})$
and $\tilde{r}_{+{\rm s}1}({\bar T})$  are
thermodynamically stable.

\begin{figure}[h]
\centering
\includegraphics[width=0.60\textwidth]{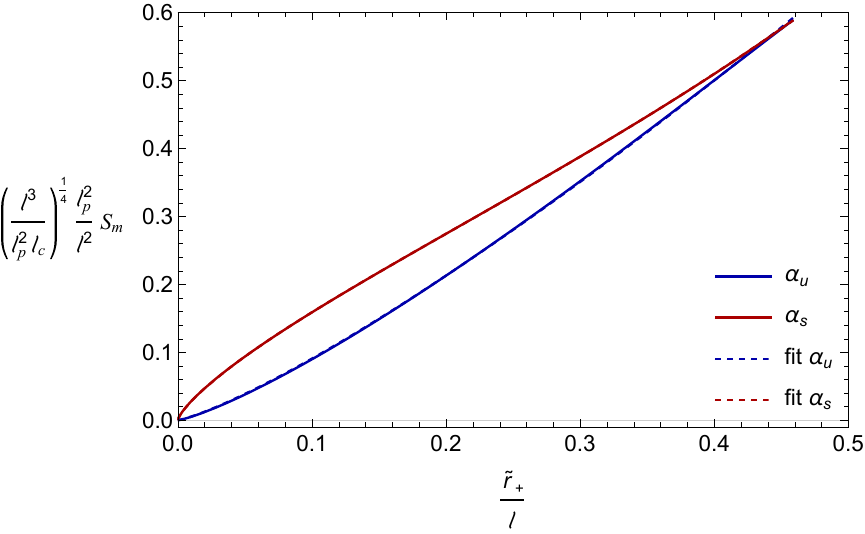}
\caption{\label{fig:entropy} Matter entropy $\left(\frac{l^3}{l_{\rm p}^2
l_{\rm c}}\right)^{\frac{1}{4}} \frac{l_{\rm p}^2}{l^2}S_{\rm m}$ in function of the
gravitational radius $\frac{\tilde{r}_+}{l}$ for the two shell radius
solutions $\alpha_{\rm u}(\tilde{r}_+)$
and $\alpha_{\rm s}(\tilde{r}_+)$.  A fit
was performed for each branch, with $\left(\frac{l^3}{l_{\rm p}^2
l_{\rm c}}\right)^{\frac{1}{4}} \frac{l_{\rm p}^2}{l^2}S_{\rm m}=1.54662
(\frac{\tilde{r}_+}{l})^{1.2323}$ for the case of
$\alpha_{\rm u}(\tilde{r}_+)$ and $\left(\frac{l^3}{l_{\rm p}^2
l_{\rm c}}\right)^{\frac{1}{4}} \frac{l_{\rm p}^2}{l^2}S_{\rm
m}=0.898912(\frac{\tilde{r}_+}{l})^{0.755675}
+0.867397(\frac{\tilde{r}_+}{l})^{2.91424}$ for
$\alpha_{\rm s}(\tilde{r}_+)$, with respective coefficients
of determination
$R^2 = 0.999992$ and $R^2=1$, with this last equality being
approximate.}
\end{figure}

The entropy $S_{\rm m}$ is now analyzed, see Fig.~\ref{fig:entropy}.
We have performed a polynomial fit to the matter entropy to understand 
its leading power of $\tilde{r}_+$.
The fit for $S_{\rm m}$
given in Eq.~\eqref{eq:entropyspecific} for the solution $\alpha_{\rm
u}$ as a function of $\tilde{r}_{+{\rm u}}$ is described by the
function $\left(\frac{l^3}{l_{\rm p}^2 l_{\rm c}}\right)^{\frac{1}{4}}
\frac{l_{\rm p}^2}{l^2}S_{\rm m}=1.54662 (\frac{\tilde{r}_{+{\rm
u}}}{l})^{1.2323}$ with a coefficient of determination $R^2=0.999992$.
The fit of $S_{\rm m}$ given in Eq.~\eqref{eq:entropyspecific} for the
solution $\alpha_{\rm s}$ as a function of $\tilde{r}_{+{\rm s}}$ is
described by the function $\left(\frac{l^3}{l_{\rm p}^2
l_{\rm c}}\right)^{\frac{1}{4}} \frac{l_{\rm p}^2}{l^2}S_{\rm
m}=0.898912(\frac{\tilde{r}_{+{\rm s}}}{l})^{0.755675}
+0.867397(\frac{\tilde{r}_{+{\rm s}}}{l})^{2.91424}$ with a
coefficient of determination $R^2=1$, with this equality being
approximate.  Another fit of $S_{\rm m}$ for the solution $\alpha_{\rm
s}$ was tried with just one power, giving $\left(\frac{l^3}{l_{\rm p}^2
l_{\rm c}}\right)^{\frac{1}{4}} \frac{l_{\rm p}^2}{l^2}S_{\rm m}=1.12374
(\frac{\tilde{r}_{+{\rm s}}}{l})^{0.867504}$, with $R^2 = 0.999607$,
however the differences of the fit are visible in the plot, and so we
have tried the fit with two powers.
These obtained fits are very near the numerical results for the
$S_{\rm m}$, which is surprising and one could wonder if there might
be an analytic solution.  But in order to obtain the expression of
$S_{\rm m}$, one needs to solve a quintic polynomial equation and we
were not able to find an analytic solution.  A feature that the fits
do not capture is the fact that $S_{\rm m}$ is only defined in the
interval $0<\frac{\tilde{r}_+}{l}<
0.4589$ with this last number being approximate.

\begin{figure}[h]
\centering
\includegraphics[width=0.60\textwidth]{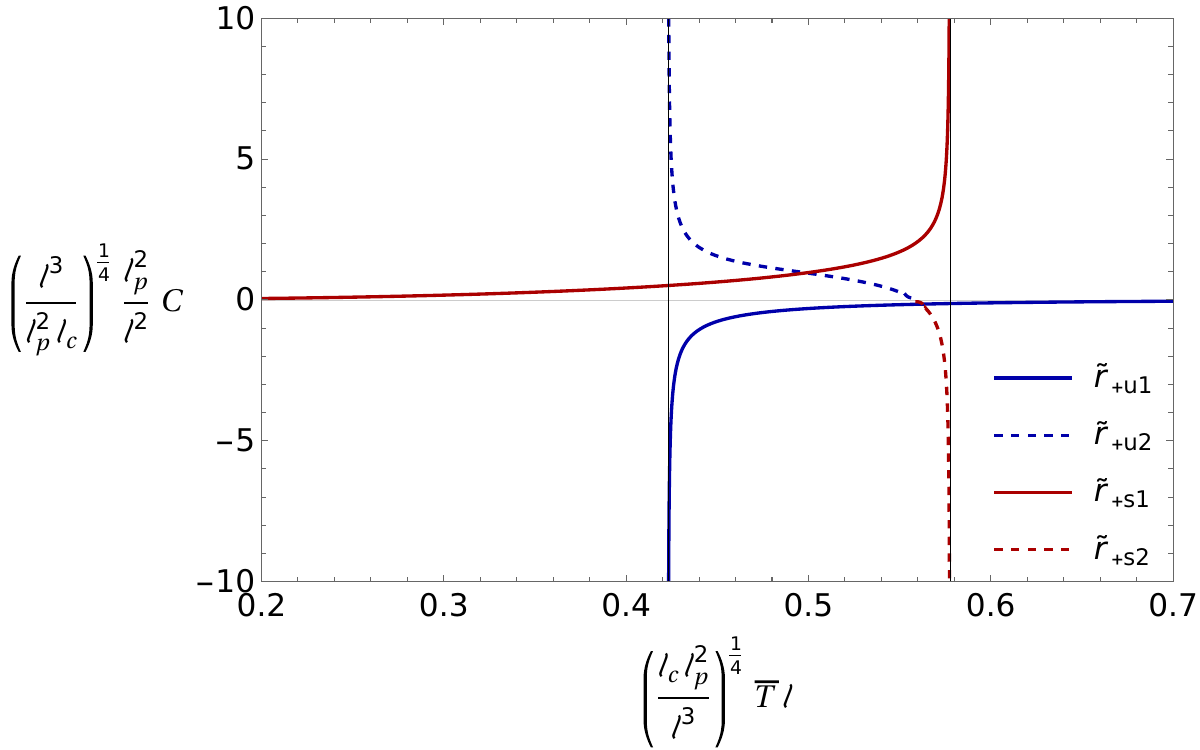}
\caption{\label{fig:Cintstar} Adimensional heat capacity for the
solutions
$\tilde{r}_{+{\rm u}1}$,
$\tilde{r}_{+{\rm u}2}$,
$\tilde{r}_{+{\rm s}1}$, and
$\tilde{r}_{+{\rm s}2}$
as functions of ${\bar T}l
\left(\frac{l_{\rm c}}{l}\right)^\frac14
\left(\frac{l_{\rm p}}{l}\right)^\frac12$,
where the solutions $\alpha_{\rm u}$
and $\alpha_{\rm s}$ are also assumed.
The solutions
$\tilde{r}_{+{\rm u}1}$ and
$\tilde{r}_{+{\rm s}2}$ are thermodynamically unstable, while
$\tilde{r}_{+{\rm u}2}$ and
$\tilde{r}_{+{\rm s}1}$ are thermodynamically stable.}
\end{figure}

Another equivalent indicator of
thermal stability is given by the positivity of the heat capacity,
see Fig.~\ref{fig:Cintstar}.  Yet, it must be noticed that
$\tilde{r}_{+{\rm u}2}({\bar T})$ has a shell radius
$\alpha_{\rm u}(\tilde{r}_{+{\rm u}2}({\bar T}))$,
which is mechanically unstable. Therefore, the only fully stable
solution is $\tilde{r}_{+{\rm s}1}({\bar T})$ with shell radius
$\alpha_{\rm s}(\tilde{r}_{+{\rm s}1}({\bar T}))$.


\section{Hot thin shell versus black hole in AdS}
\label{tsxbh}

\subsection{The black hole}

For completeness,
we now give the relevant quantities of the canonical 
ensemble of a Schwarzschild-AdS black hole, which are
important for the comparison between
a hot thin shell and a black hole in AdS. 
In order to obtain the reduced 
action and then the zero-loop
action, one can carry on in a similar way the 
calculations above but now with black hole boundary conditions rather than
thin shell ones, or one can simply use the Hawking-Page results.

The
action of the Hawking-Page black hole solutions is
\begin{align}
I_{\rm bh} = 
\frac{1}{2 l_{\rm p}^2{\bar T}}
\left( r_+ + \frac{r_+^3}{l^2}
\right)  - \frac{\pi r_+^2}{l_{\rm p}^2}
\,,
\label{eq:actionbh}
\end{align}
with the radius $r_+$ being a function of ${\bar T}$.
Indeed here, the gravitational radius $r_+$, which
is also a horizon radius since we are now dealing
with a black hole, is
given by the equation
${\bar\beta} \equiv\frac{1}{{\bar T}}= \frac{\iota(r_+)}
{1 + 3\frac{r_+^2}{l^2}}$, see also Eq.~\eqref{eq:statio2final} with
$\iota(r_+)=4\pi r_+$.

\begin{figure}[h]
\centering
\includegraphics[width=0.60\textwidth]{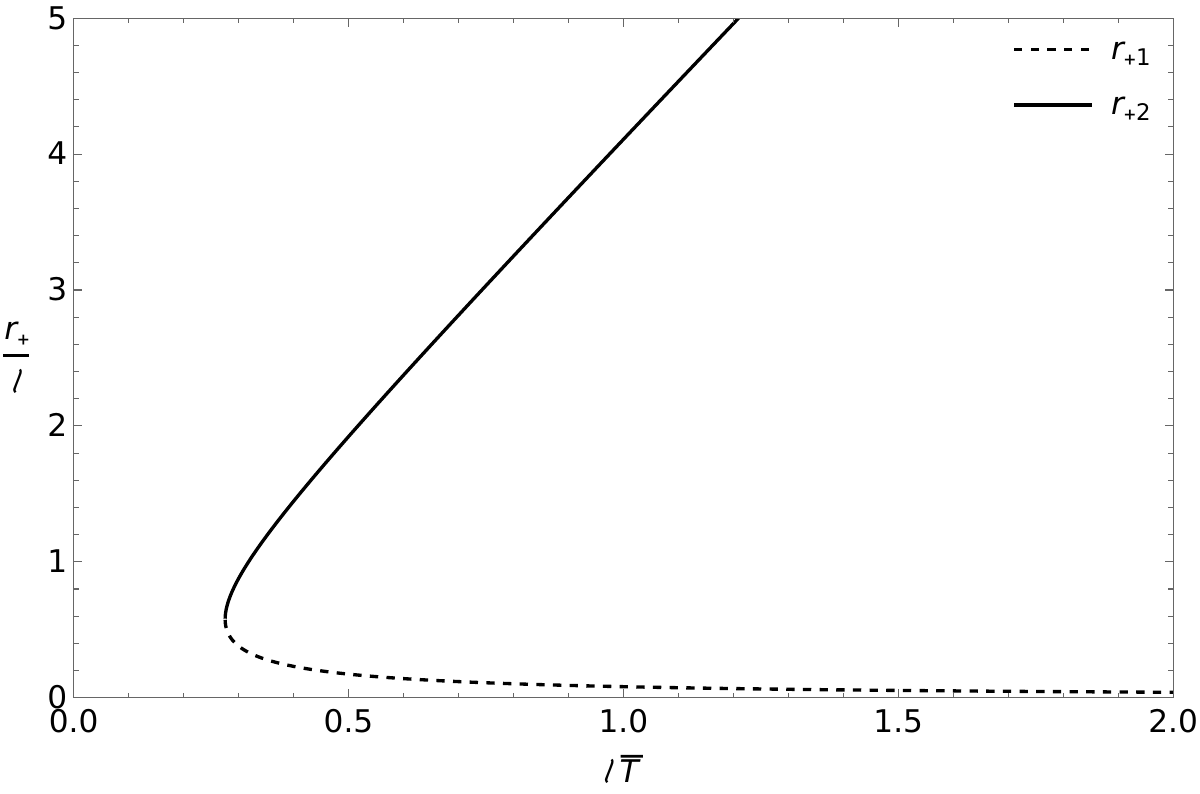}
\caption{\label{fig:r+bh} Solutions of the ensemble 
$\frac{r_{+1}}{l}$
and 
$\frac{r_{+2}}{l}$
for the black hole in asymptotically AdS.}
\end{figure}

Therefore, one has to solve  for $r_+$ the equation
${\bar T}=\frac{1 + 3\frac{r_+^2}{l^2}}{4\pi r_+}$.
The solutions
$\frac{r_+}{l}$ as a function of ${\bar T}l$ are given
by
\begin{align}
\frac{r_+}{l} = 
\frac{2\pi l {\bar T}}{3} \pm \frac{1}{3}\sqrt{(2\pi l {\bar T})^2 - 3}.
\label{eq:radiusbh}
\end{align}
Thus, for $l {\bar T}\geq\frac{\sqrt3}{2\pi}$, there are two black
hole solutions, $r_{+1}({\bar T})$, the solution with
the minus sign, which is thermodynamically unstable, and $r_{+2}({\bar T})$,
the solution with the plus sign, 
which is stable. 
When equality holds, $l {\bar T} = \frac{\sqrt3}{2\pi}$, one has a degenerate
solution,
$r_{+1}({\bar T})=r_{+2}({\bar T})=\frac{1}{\sqrt{3}}$.
For $l {\bar T}<\frac{\sqrt3}{2\pi}$,
there are no solutions, see Fig.~\ref{fig:r+bh}. 
The numerical value of this 
critical temperature is
$l {\bar T}=\frac{\sqrt3}{2\pi}=0.276$ approximately, with 
the corresponding horizon radius $\frac{r_+}{l}=\frac{1}{\sqrt{3}}=0.577$.

\begin{figure}[h]
\centering
\includegraphics[width=0.60\textwidth]{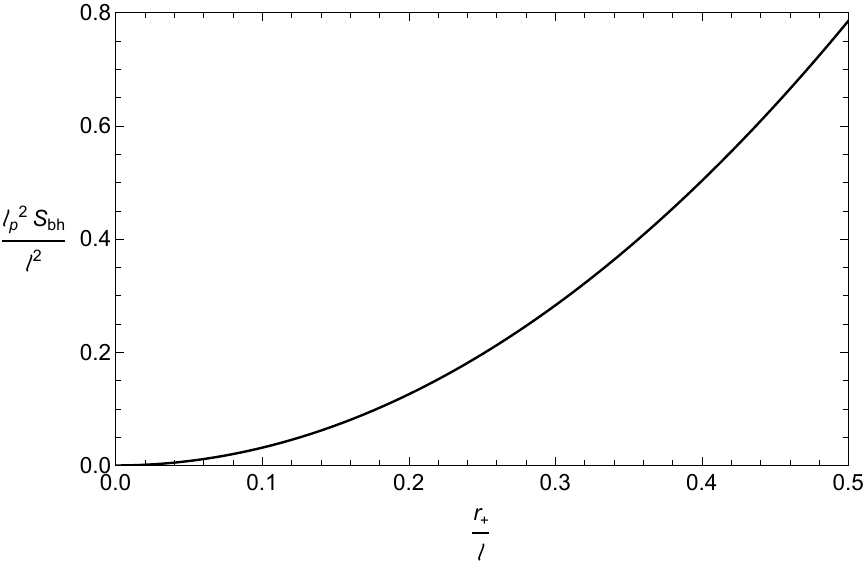}
\caption{\label{fig:entropybh} Black hole entropy 
$\frac{l^2_{\rm p}}{l^2}S_{\rm bh}$ as a function of 
the horizon radius $\frac{r_+}{l}$,
which stands either for $\frac{r_{+1}}{l}$
or for $\frac{r_{+2}}{l}$.}
\end{figure}

The black hole entropy can be obtained from the action as
\begin{align}
S_{\rm bh}=\frac{\pi r_+^2}{l_{\rm p}^2}\,,
\label{eq:Sbh}
\end{align}
i.e., the Bekenstein-Hawking entropy, with
$r_+$ standing for $r_{+1}$ or $r_{+2}$. 
The entropy describes the usual parabola, see
Fig.~\eqref{fig:entropybh}.

\begin{figure}[h]
\centering
\includegraphics[width=0.60\textwidth]{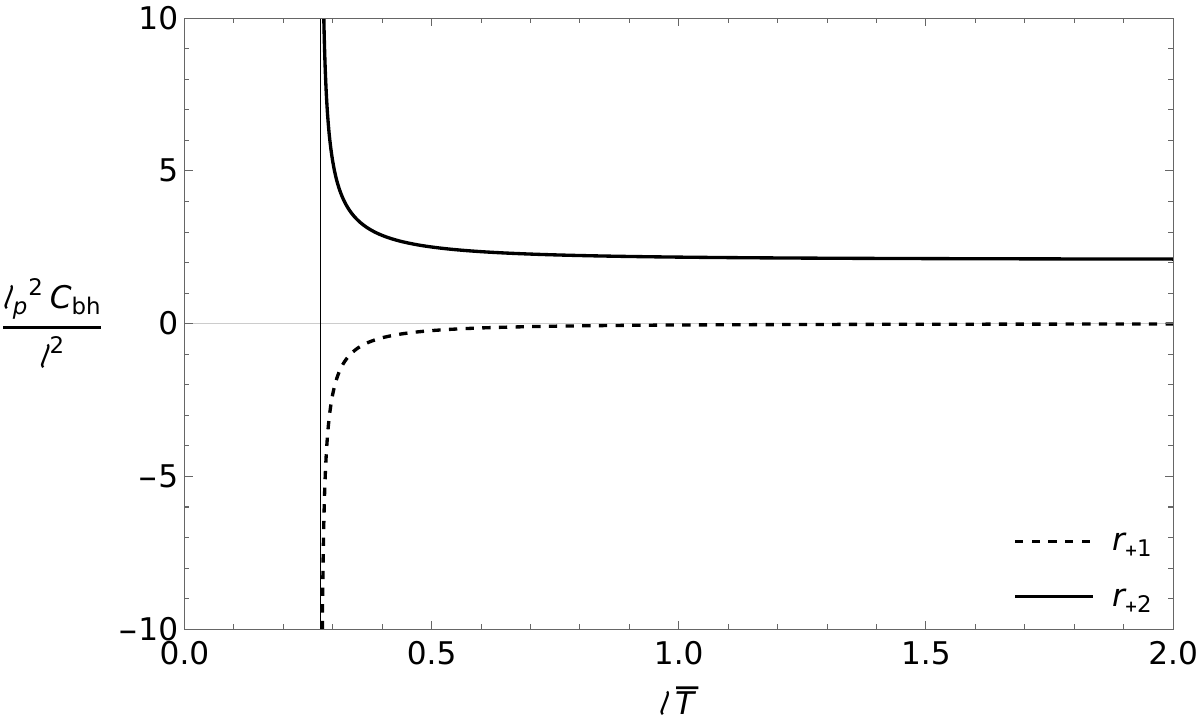}
\caption{\label{fig:Cbh} Adimensional heat capacity $\frac{l_{\rm p}^2
C_{\rm bh}}{l^2}$ of the black hole solutions $r_{+1}$ and $r_{+2}$ in
function of ${\bar T} l$.}
\end{figure}

The heat capacity for the Schwarzschild-AdS black hole is 
\begin{align}
C_{\rm bh}=\frac{2\pi r_+^2 \left(1 + 3\frac{\tilde{r}_+^2}{l^2}\right)}
{3\left(\frac{r_+^2}{l^2} -
\frac{1}{3}\right)}\,, \label{eq:Cbh}
\end{align}
for each solution $r_{+1}({\bar T})$ and $r_{+2}({\bar T})$, see
Fig.~\ref{fig:Cbh}.  The heat capacity is positive for $\frac{r_+}{l}
> \frac{\sqrt{3}}{3}$, and so $r_{+1}$ is thermodynamic unstable, and
$r_{+2}$ is thermodynamic stable.

It is of relevance to remark that pure hot AdS and
black hole in AdS compete to be the prominent
thermodynamic phase, with the phase that has the minimum action
being the one that is favored. Pure hot AdS has zero action,
$I_{\rm PAdS}=0$, so if $I_{\rm bh}>0$ then AdS is favored,
if $I_{\rm bh}=0$ the two phases coexist equally,
and if $I_{\rm bh}<0$ then the black hole is favored. 
From the black hole action, Eq.~\eqref{eq:actionbh}, 
one finds that
as one increases $l {\bar T}$ from zero, there is a
first order phase transition
from thermal AdS to the stable black hole state, the transition
happening at $l {\bar T}=\frac{1}{\pi}=0.318$, the latter
number being approximate, to the stable black 
hole with horizon radius $\frac{{r}_{+2}}{l}=1$.

\subsection{Hot thin shell versus black hole and favorable states}

\subsubsection{Gravitational radii, entropies, and heat capacities}

It is now of interest to compare the properties we have obtained for
the hot
thin shell in AdS with the properties of the black hole in AdS.
More specifically, we can compare the gravitational radii of each thin
shell with the gravitational radius of the black hole, examine their
entropies, and analyze their heat capacities.

Let us start with the comparison of the two possible gravitational
radii of each thin shell with the gravitational radii of the black
hole.
For the mechanically unstable hot thin shell, which is given by the shell
radius $\alpha_{\rm u}$, we have found that there are two branches for
the gravitational radius. One of the branches, $\tilde{r}_{+{\rm u}1}$
is thermodynamically unstable, while the other branch
$\tilde{r}_{+{\rm u}2}$ is thermodynamically stable.  As well, for the
Hawking-Page black hole, the horizon radius $r_+$ has one branch
$r_{+1}$, which is thermodynamically unstable, and another branch
$r_{+2}$, which is thermodynamically stable.  It is clear from
Fig.~\ref{fig:alphainrp} and Fig.~\ref{fig:r+bh} that the two
solutions for the gravitational radius with thin shell radius
$\alpha_{\rm u}$, which is mechanically unstable, share similarities
with the two solutions for the horizon radius of the black hole. In
detail, the thermodynamically unstable branch of the thin shell
solution, $\tilde{r}_{+{\rm u}1}$, follows the same behavior as the
thermodynamically unstable black hole solution, $r_{+1}$. As well, the
thermodynamically stable branch of the thin shell solution,
$\tilde{r}_{+{\rm u}2}$, follows the same behavior as the
thermodynamically stable black hole solution, $r_{+2}$.  These
similarities could perhaps be expected since the mechanically unstable
shell is bound to collapse into a black hole and can be understood as
a black hole precursor.
For the mechanically stable hot thin shell, which is given by the shell
radius $\alpha_{\rm s}$, we have found that there are also two
branches for the gravitational radius. One of the branches,
$\tilde{r}_{+{\rm s}1}$ is thermodynamically unstable, while the other
branch $\tilde{r}_{+{\rm s}2}$ is thermodynamically stable.
From Fig.~\ref{fig:alphainrp} and Fig.~\ref{fig:r+bh}, it can be 
seen that the
two solutions for the gravitational radius with thin shell radius
$\alpha_{\rm s}$, which is mechanically stable, share no similarities
with the two solutions for the horizon radius of the black hole.
However, the solutions $\tilde{r}_{+{\rm s}1}$ 
and $\tilde{r}_{+{\rm s}2}$ appear to have similarities with the
behavior of the Davies black hole solutions, which correspond to an
electrically charged black hole \cite{fernandeslemos2024} in the
canonical ensemble. These black hole solutions have a stable branch
that start at zero temperature with a horizon radius given by the
electric charge, and then the horizon radius increases with the
temperature up until a maximum temperature.  The same happens with the
mechanically and thermodynamically stable hot
matter thin shell, starting
at zero temperature with zero gravitational radius instead of a
non-zero value. This behavior is expected from a solution of hot
self-gravitating matter that models hot AdS space
with radiation at the same order of approximation.

Let us now compare the matter thin shell entropy with the black hole
entropy. Both entropies depend on their own gravitational radius,
which in the black hole case is also a horizon radius.
For the mechanical unstable and stable thin shells we have seen that
the matter entropy $S_{\rm m}$
can be described approximately by a power law.
For the unstable shell $\alpha_{\rm u}$, we found that $S_{\rm m} = \xi
\tilde{r}_{+}^{\;1.2323}$, for some $\xi$.
For the stable shell $\alpha_{\rm s}$, we
found that $S_{\rm m} = \xi \tilde{r}_{+}^{\;0.867504}$, for some
other $\xi$.
For the black hole, the entropy is also given by a power law $S_{\rm
bh}= \chi r_+^2$, for some $\chi$.
We see that both the mechanical unstable shell $\alpha_{\rm u}$ and
the black hole have exponents $\gamma $ satisfying $\gamma >1$,
while the stable
shell has an exponent $\gamma <1$. This behavior reinforces the similarity
properties between the mechanical unstable shell and the black hole,
as advocated above. To understand this we resort
to a Bekenstein argument \cite{Bekenstein:1973}.
It is stated in it, without making calculations, that one should
expect that black holes have an entropy with an exponent $\gamma $
obeying $\gamma >1$. 
Suppose that they had an exponent $\gamma <1$, then 
one would have that two isolated black holes that will merge into one
should have a final mass lower than the sum of the initial
mass, part of the initial mass being lost in
gravitational radiation. Concomitantly
the entropy of the final black hole
should be greater than the sum of the
entropies of the initial black hole, to have the second
law of thermodynamics obeyed.
The two conditions cannot be met simultaneously when $\gamma <1$.
Suppose for these purposes that the black hole
entropy is proportional to a power of the gravitational radius,
$S_{\rm bh}= \chi r_+^\gamma$,
for some $\chi$ and $\gamma$.
For black hole $a$ and black hole $b$ merging into 
a third black hole $c$,
one has from the first condition $r_{+c}< r_{+a} + r_{+b}$ and from 
the second condition $r_{+c}^\gamma > r_{+a}^\gamma + r_{+b}^\gamma$, i.e., 
one has the range for the horizon radius of the black hole $c$
obeying
$(r_{+a}^\gamma + r_{+b}^\gamma)^{\frac{1}{\gamma}} < r_{+c} < 
r_{+a} + r_{+b}$.
This inequality can only be fulfilled for $\gamma>1$. 
From thermodynamic arguments then it was chosen $\gamma=2$,
the correct value for black holes.
The point here we want to make is that
the unstable shell $\alpha_{\rm u}$ has an exponent
$\gamma=1.2323$ that is indeed greater than one, and
following the arguments above it is black hole like, i.e.,
the shell behaves as the black hole that it can originate
from gravitational collapse.
The 
stable shell $\alpha_{\rm s}$ has an exponent
$\gamma=0.867504$ that is less than one, and
thus it does not behave as a black hole that it could
originate upon collapse.
Another remark
in relation to the entropy
is that for the black hole $S_{\rm bh}= \pi
\frac{l^2}{l_{\rm p}^2}\left(\frac{r_+}{l}\right)^2$,
the exponent is $\gamma=2$,
and it grows much faster than for the matter for large gravitational
radius, as expected, since it is known that black holes have the
maximum entropy.  However, for small gravitational radius, the matter
entropy $S_{\rm m}$ is larger than the black
hole entropy, so this might indicate that there are no
stable
black holes for small gravitational radius, see Fig.~\ref{fig:r+bh}
noting that $r_{+1}$ is unstable.

Let us continue with the comparison of the heat capacities for the
hot thin
shell and black hole, $C$ and $C_{\rm bh}$, respectively, with the
help of Figs.~\ref{fig:Cintstar} and \ref{fig:Cbh}.
The heat capacity $C$ as a function of the temperature for the
mechanically unstable shell $\alpha_{\rm u}(\tilde{r}_+)$ behaves in
the same manner as the heat capacity of the black hole $C_{\rm bh}$ as
a function of temperature. There are parts that are thermodynamically
unstable, $\tilde{r}_{+{\rm u}1}$ for the shell and $r_{+1}$ for the
black hole, and parts that are thermodynamically stable
$\tilde{r}_{+{\rm u}2}$ for the shell, and $r_{+2}$ for the black
hole.
The heat capacity $C$ as a function of the temperature for the
mechanically stable shell $\alpha_{\rm u}$ behaves differently from the
heat capacity of the black hole $C_{\rm bh}$ as a function of
temperature. However, it shows similarities to the heat capacity of
the electrically charged black hole in the canonical
ensemble~\cite{fernandeslemos2024}.  In particular, the
thermodynamically stable $\tilde{r}_{+{\rm s}1}$ branch of the heat
capacity is similar to the heat capacity of the stable branch of
an electrically charged black hole.  These similarities
displayed here for
the heat capacity, are the same as the similarities we have found
above when
comparing the gravitational radii, and indeed they come from the fact
that the heat capacity is related to the first derivative in
temperature of the gravitational radius, and so the similarities from
the gravitational radius solutions are carried into the heat capacity.

\vfill

\subsubsection{Favorable states: Comparison between hot thin shell,
black hole, and pure hot AdS thermodynamic states}

We want now to identify the favorable states of the ensemble, i.e.,
given a fixed temperature we want to know if the
hot thin shell is favored in relation to the black hole
or if the contrary happens.

In order to do this,
to identify the favorable states at a fixed temperature 
of the ensemble, we must compare the action of the stable hot
self-gravitating matter thin shell with the action of the stable black
hole solution of Schwarzschild-AdS. This is so because the sector with
a self-gravitating matter shell may compete with the black hole sector
and the pure hot AdS sector in the path integral.
From thermodynamics, it is known that the preferred
configuration is the one with the least free energy.
This also means the one with the least
action, since $I_0= {\bar\beta} F$.
This can be seen because the partition function is $Z={\rm e}^{-I_0}$,
and thus, the configuration with less $I_0$ is the more probable one.
Now, if at a certain temperature, the
configuration with the least action changes, then this marks a first order
phase transition as the action is continuous but not differentiable
there.  In~\cite{hawkingpage1983}, hot thermal AdS, i.e.,
hot AdS in one-loop approximation, and the stable
black hole solution were discussed, where it was discovered the
Hawking-Page phase transition from hot thermal AdS to the stable black
hole. Here, the stable self-gravitating matter thin shell can be 
understood as one possible description of hot AdS with thermal 
self-gravitating matter, i.e., hot curved AdS.

To help in the comparison of the actions, 
one can write the action of the matter thin shell
given in Eq.~\eqref{eq:reducedactionpathintegral} as
$\frac{l_{\rm p}^2}{l^2}I_0 =
\frac{\frac{\tilde{r}_+}{l} \left( 1 + \frac{\tilde{r}_+^2}{l^2}
\right)}{2 {\bar T}l} -
c_0\left( \frac{l_{\rm p}^2l_{\rm c}}{l^3} \right)^\frac14
\left(\frac{m \,l_{\rm p}^2}{l}\right)^{\frac{3}{4}}
\left(4\pi \frac{\alpha^2}{l^2}\right)^{\frac{1}{4}}$,
with $\tilde{r}_+$
given by
the stable solution
$\tilde{r}_+=\tilde{r}_+({\bar T})$
of Eq.~\eqref{eq:statio2final}, i.e.,
$\tilde{r}_{+{\rm s}2}$,
and $\alpha$ given by 
the stable solution
$\alpha=\alpha(\tilde{r}_+({\bar T}))$ of Eq.~\eqref{eq:statio1final}
together
with Eq.~\eqref{eq:statio2final}, i.e.,
$\alpha_{\rm s}=\alpha_{\rm s}(\tilde{r}_{+{\rm s}2}({\bar T}))$,
and the action has been set without units.  It is useful to define
$z=\left(\frac{l^3}{l_{\rm p}^2l_{\rm c}} \right)^\frac14$ as the
parameter without units
that establishes the relevant scale ratios between the
Planck length, AdS length, and the Compton length in the case of the
hot matter thin shell with the chosen equations of state above.
Since the parameter $z$ is given by the ratio between the 
scales of the matter entropy and the black hole
entropy, 
its value establishes how dominant the matter entropy is compared to the 
black hole
entropy, which affects the phase structure, as we 
shall see.
Then, one can write the
unitless action of the matter thin shell as
\begin{align}
&
\frac{l_{\rm p}^2}{l^2}I_0=
\frac{l_{\rm p}^2}{l^2}
\frac1z {\bar
I}_0\left(\frac{{\bar T} l}{z}\right),
\quad\quad\quad\quad\quad\quad
z\equiv\left(\frac{l^3}{l_{\rm p}^2l_{\rm c}} \right)^\frac14
\,,
\nonumber
\\
&\frac{l_{\rm p}^2}{l^2}{\bar I}_0\left(\frac{{\bar T}
l}{z}\right)\equiv 
\frac{\frac{\tilde{r}_+}{l} \left( 1 +
\frac{\tilde{r}_+^2}{l^2} \right)}{2 \frac{{\bar T}l}{z}} -
 c_0 \left(\frac{m
\,l_{\rm p}^2}{l}\right)^{\frac{3}{4}}
\left(4\pi \frac{\alpha^2}{l^2}\right)^{\frac{1}{4}}\,.
\label{eq:Inounits}
\end{align}
This property of the thin shell action
is also useful for numerical purposes, as one can compare
actions 
with a given parameter $z$.
The behavior of the action of the
matter thin shell, Eq.~\eqref{eq:Inounits}, evaluated at the 
solutions of the shell, can be summarized as follows. 
The action is zero
at $l{\bar T}=0$, and it decreases for increasing $l{\bar T}$. 
At a final temperature $l\bar{T}_{\rm f}$, the stable thin shell solution 
ceases to exist, which can be
interpreted as
the matter having larger thermal agitation than the permitted to
have a shell, implying that the shell can collapse to a black
hole or disperse to infinity. This $l\bar{T}_{\rm f}$ depends on the 
parameter
$z
=
\left( \frac{l^3}{l_{\rm p}^2l_{\rm c}} \right)^\frac14$,
which itself depends
on the natural gravitational scale ratio $\frac{l}{l_{\rm p}}$ and on the
matter scale ratio $\frac{l}{l_{\rm c}}$.

In relation to the action of the stable black hole, one has
the Hawking-Page action given by
\begin{align}
\frac{l_{\rm p}^2}{l^2}I_{\rm bh} = 
\frac{\frac{r_+}{l} \left( 1 + \frac{r_+^2}{l^2}
\right)}{2 {\bar T}l}  - 
\frac{\pi r_+^2}{l^2}
\,,
\label{eq:actionbhrepeat}
\end{align}
with $r_+$ being the stable black hole
solution, i.e.,
$r_{+2}$, 
given
as
a function of $l{\bar T}$
by
$\frac{\tilde{r}_+}{l} = 
\frac{2\pi l {\bar T}}{3} + \frac{1}{3}\sqrt{(2\pi l {\bar T})^2 - 3}$.
We have seen that the stable solution only exists
for $l {\bar T}\geq\frac{\sqrt3}{2\pi} =0.276$, with last equality 
being approximate.

\begin{figure}[h]
\centering
\includegraphics[width=0.60\textwidth]{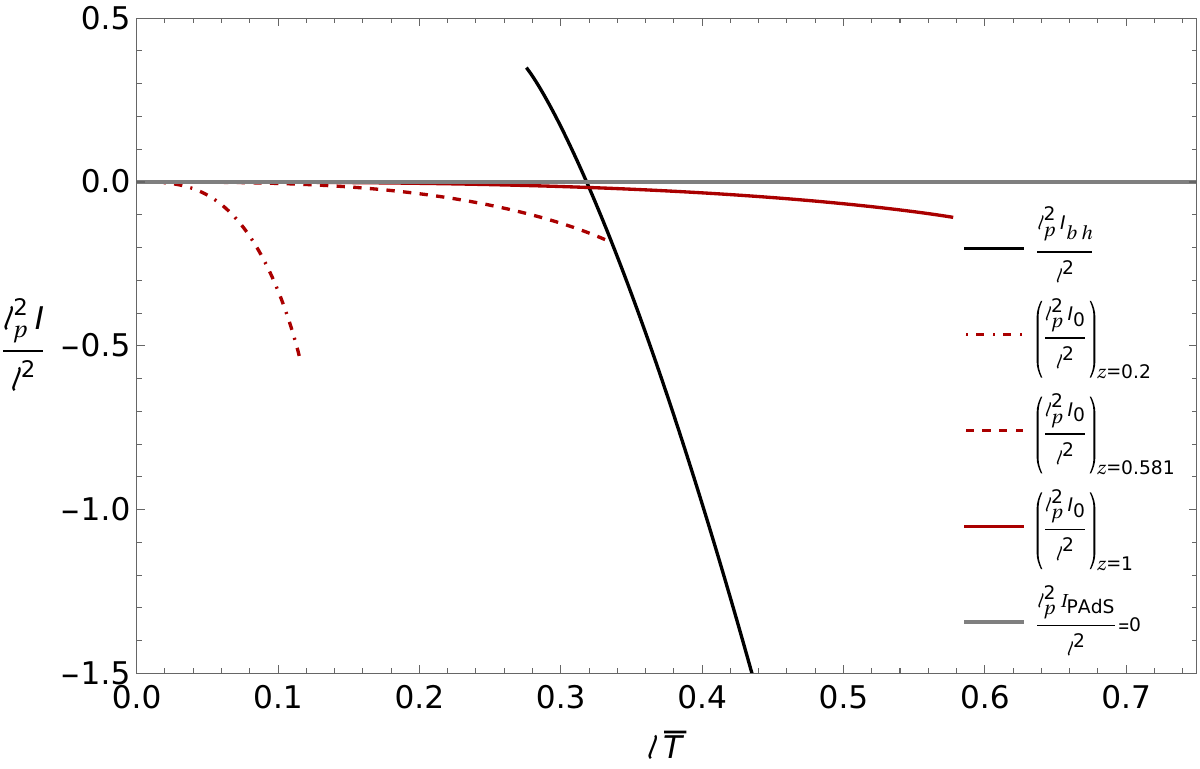}
\caption{\label{fig:actionintstar}
Plot of the actions
$I_{\rm bh}$, $I_0$, and 
$I_{\rm PAdS}$
as functions of the temperature $l {\bar T}$. The
solution that has lower action between stable black
hole, hot shell, and pure hot AdS
is the one that is favored.  It is chosen
$z=\left( \frac{l}{l_{\rm c}}\right)^\frac{1}{4} \left(\frac{l}{l_{\rm
p}}\right)^\frac{1}{2} = 0.2,\, 0.581,\,1$ to compare the actions.
For $z=0.2$, the hot shell ceases to exist at temperature 
$l \bar{T}_{\rm f} = 0.115$, for
$z=0.581$ at temperature 
$l \bar{T}_{\rm f} = 0.335$, and for 
$z=1$, at temperature 
$l \bar{T}_{\rm f}=0.577$.
}
\end{figure}

Now, we must compare the matter action with the black hole action for
each possible parameter $z$ and $c_0$, and also with
pure hot AdS space
characterized by $I_{\rm PAdS} = 0$.  Here, we make the choice $c_0=1$
since $c_0$ can be in some sense absorbed by the parameter $z$, and
then make the comparison of the actions only on $z$.  The plot of the
actions is shown in Fig.~\ref{fig:actionintstar}, where for the matter
shell we put three values of $z$,
namely, $z =0.1$, $z=0.581$, and $z=1$.  
The action of matter thin shell 
starts from zero at $l\bar{T}=0$ and decreases until 
a minimum negative value for a maximum temperature $l\bar{T}_{\rm f}$. 
Note that the maximum temperature depends on the parameter 
$z$ as $l \bar{T}_{\rm f} = 0.577 z$, where $0.577$ is approximate.
Above this temperature $l \bar{T}_{\rm f}$ the shell stops
to exist and probably collapses.
%
With respect to the black hole case, the action $I_{\rm bh}$ 
only starts to exist for $l {\bar T}\geq \frac{\sqrt3}{2\pi}=0.276$,
and decreases with increasing
temperature. The black hole  action is positive in the 
range $\frac{\sqrt3}{2\pi} \leq l \bar{T} \leq 0.318$
where $0.318$ is approximate, is zero at $l\bar{T} = 0.318$,
and is negative for $l {\bar T}>0.318$.
The point of intersection between the two actions is given by the
equality $I_{\rm bh}(l \bar{T}) = \frac{1}{z} \bar{I}_0(\frac{l
\bar{T}}{z})$ for each $z$. For example, in the case of $z=1$, one has
that the actions intersect at $l {\bar T} = 0.320$, and so as one increases
the temperature around this point, there is a first order phase
transition from the hot
matter thin shell to the stable black hole.  This
is analogous to the case of the Hawking-Page phase transition, where
the matter is treated in AdS space
at one-loop approximation, rather than in
zero loop.
It can be found that the intersection between the matter thin shell
action and the black hole action only happens for a range of $z$.  As
one decreases $z$, the maximum temperature of the thin shell also
decreases, while the black hole action is unaltered. And so, there
must be a minimum value of $z$ for which the intersection occurs. The
minimum value can be found by considering that the two actions
intersect exactly at the maximum temperature of the shell,
i.e., $I_{\rm bh}( 0.577 z) = \frac{1}{z} \bar{I}_0( 0.577)$, which
numerically can be solved and gives $z= 0.581$ approximately and the
first order phase transition occurs for this case at $l {\bar T}=0.336$,
approximately. As a consequence, the action of the matter thin shell
intersects the action of the black hole only if
$0.581\leq z <\infty$, with
first number being approximate. If $0< z < 0.581$, the matter shell
solution ceases to exist before it intersects the curve of the action
of the black hole. Therefore, there is only a first order phase
transition from the matter thin shell to the black hole when
$0.581\leq z <\infty$.

A comment can be made regarding pure hot AdS space,
i.e., AdS space in zero loop,
in comparison with the hot thin shell and
black hole configurations. Since the matter action is always negative, 
the matter thin shell is always more favorable than pure hot AdS space 
as long as the thin shell solution exists. However, we must also consider 
the existence of the stable black hole. For $0.581<z<\infty$, 
the matter thin shell action always intersects
the black hole action and so hot pure AdS is never the favorable phase.
For
$0.547 \leq z \leq 0.581$, the matter thin shell and the black hole 
actions do not intersect but at the point where the matter thin shell 
ceases to exist, the action of the black hole is negative,
and so these configurations are more favorable than pure hot AdS space. 
For the range $0\leq z \leq 0.547$, the matter thin shell ceases to exist 
at a temperature at which either there is no black hole solution or 
the action of the black hole solution is positive, so there is an 
interval of temperatures where the pure hot AdS space is favorable.

It is worth stressing that
the parameter $z$ can be restricted from validity arguments of the
zero-loop approximation. The zero-loop approximation should be valid
for the cases $l \gg l_{\rm p}$ but with $l$ not that large and since
$\alpha$ is comparable to $l$, one must have $l \gg l_{\rm c}$ also so that
matter at the thin shell can be judged thermodynamic.
Moreover, all scales must be much greater than the Planck scale
$l_{\rm p}$, as we are using the zero-loop approximation.  Therefore,
we must have $l \gg l_{\rm c} \gg l_{\rm p}$, which means a large
value of $z=\left( \frac{l^3}{l_{\rm p}^2l_{\rm c}}
\right)^\frac14$. In this regime, the first order phase transition can
always occur, and both the matter thin shell and the black hole
solutions are more favorable than pure hot AdS space $I_{\rm PAdS}=0$.
This strengthens the interpretation that the matter thin shell with
the chosen equation of state models hot AdS space with
self-gravitating radiation matter at low temperatures.

\subsubsection{Favorable states: Comparison between thin shell, black
hole, and hot thermal AdS thermodynamic states}

We have just seen how the action
of a stable self-gravitating matter system in AdS,
which is a realization of hot curved AdS, compares with
the action of the stable AdS black hole solution,
and the action of pure AdS, describing classical
AdS space devoid of any matter.

It is also interesting to substitute pure AdS for hot thermal AdS,
i.e., AdS space with
nonself-gravitating radiation obtained from the one-loop approximation, 
and compare with the action for black
hole and the thin shell. 
The action $I_{\rm TAdS}$ for hot thermal AdS is given by
$I_{\rm TAdS}=-\frac{\pi^4 (l {\bar T})^3}{45}$,
where we have considered that  hot thermal AdS is made of
particles, each with 
effective number of spin states equal to $2$, such as gravitons do.
For the
purpose of comparison with the other
actions we write $I_{\rm TAdS}$ as
\begin{align}
\frac{l_{\rm p}^2}{l^2}I_{\rm TAdS}=-\frac{l_{\rm p}^2}{l^2}
\frac{\pi^4 (l {\bar T})^3}{45}\,.
\label{eq:actionhotthermalads}
\end{align}

\begin{figure}[h]
\centering
\includegraphics[width=0.60\textwidth]{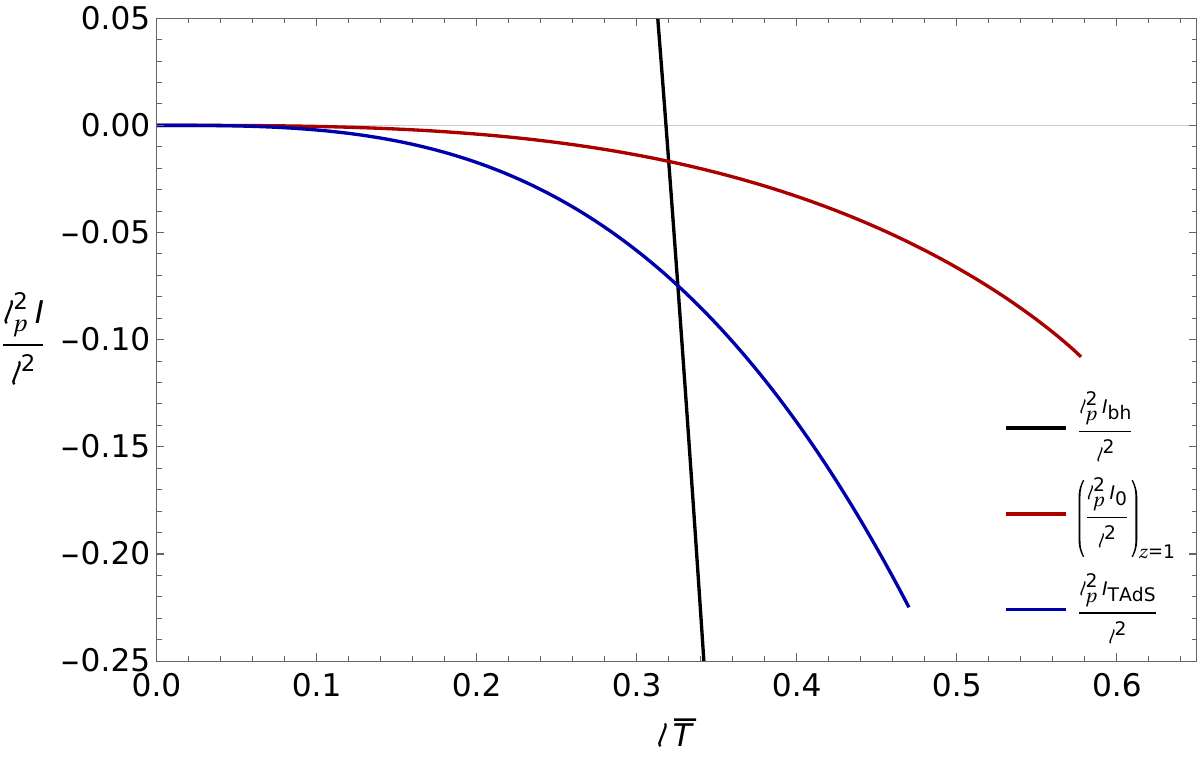}
\caption{\label{fig:actionradiation}
Plot of the actions
$I_{\rm bh}$, $I_0$, and 
$I_{\rm TAdS}$ as functions of the temperature $l {\bar T}$. The
solution that has lower action between  stable black
hole, hot shell, and thermal hot AdS,
i.e., AdS with nonself-gravitating radiation,
is the one that is favored.
It is chosen $z=\left( \frac{l}{l_{\rm c}}\right)^\frac{1}{4} 
\left(\frac{l}{l_{\rm p}}\right)^\frac{1}{2} = 1$ and 
$\frac{l_{\rm p}}{l} = 1$ to compare the actions.
For $z=1$, the hot shell ceases to exist at temperature 
$l \bar{T}_{\rm f} = 0.577$. 
Thermal hot AdS ceases to exist at temperature
$lT_{\rm Buch}=0.4701$.}
\end{figure}

We can then compare the first
order phase transition treated above
with the Hawking-Page phase
transition, which is a transition between hot thermal AdS
with action given in  Eq.~\eqref{eq:actionhotthermalads}
and the black hole
action given in Eq.~\eqref{eq:actionbh} \cite{hawkingpage1983}.
Moreover, when the temperature
of the radiation is sufficiently high, 
the radiation forms a singularity of the Buchdahl type
in the center and then it presumably collapses
to a black hole. One finds that 
this maximum Buchdahl radiation temperature
for which radiation ceases to exist is given by 
$lT_{\rm Buch}=0.4234\left(\frac{2\pi^2}{30}\right)^{-\frac14}
\left(\frac{l}{l_{\rm p}}\right)^{\frac12}$
\cite{hawkingpage1983}, see also \cite{pagephillips:1985}.
In Fig.~\ref{fig:actionradiation}, the actions are plotted for
the stable black hole, the
hot matter thin shell, and hot thermal AdS.
In the figure, we have used the numbers
$z=\left( \frac{l}{l_{\rm c}}\right)^\frac{1}{4} 
\left(\frac{l}{l_{\rm p}}\right)^\frac{1}{2} = 1$ and 
$\frac{l_{\rm p}}{l} = 1$ to compare the actions.
First, we comment on the transitions
between stable black hole and hot thermal AdS thermodynamic states,
and second, we compare the results for the
hot matter thin shell and hot thermal AdS.
In relation to the first point, we
have seen that there are no black holes, and therefore,
no stable black hole, in the range of temperatures
$0\leq l{\bar T}<\frac{\sqrt3}{2\pi}=0.276$, the last number
being approximate.
From 
$\frac{\sqrt3}{2\pi}\leq l{\bar T}<0.325$, hot thermal AdS
is favored in relation to a black hole state.
At $l{\bar T}=0.325$ hot thermal AdS and black hole coexist equally,
and this is the temperature at which a first order
phase transition occurs. 
This is the Hawking-Page phase transition.
For
$0.325<l{\bar T}<lT_{\rm Buch}$, where $lT_{\rm Buch}=0.4701$, the black
hole is favored over hot thermal AdS, meaning that is more probable to
find the system in a black hole state.  For $lT_{\rm
Buch}<l{\bar T}<\infty$, the system is in a collapsed black hole state in
AdS, meaning that at these temperatures, it is not possible to find the
system in a hot thermal AdS state.
In relation to the second point, we see from the figure 
that the first order phase transition between the matter thin
shell and the black hole is analogous to the Hawking-Page
phase transition between hot thermal AdS and the black hole.
However,
it seems that  hot thermal AdS, i.e., AdS with nonself-gravitating
radiation, is more favorable than the matter thin shell, with the
differences being very small as one increases $z$ and $\frac{l}{l_{\rm p}}$.
Since the action for hot thermal AdS does not include gravitation, as
it corresponds to nonself-gravitating radiation, it is clear that the
hot thin shell should mimic self-gravitating radiation due to its
analogous behavior around the phase transition.
In addition, we see they share the feature of a maximum temperature,
$l \bar{T}_{\rm f}= 0.577$
for the hot thin shell and
$lT_{\rm Buch}=0.4701$
for hot thermal AdS, the values being for $z=1$ and $l=l_{\rm p}$.

\section{Conclusions}
\label{sec:concl}

In this work, we have studied the canonical ensemble of a hot
self-gravitating matter thin shell in AdS by finding the partition
function of the system via the Euclidean path integral approach to
quantum gravity. We have restricted the study to spherically symmetric
metrics and have established the boundary conditions. Imposing the
Hamiltonian constraint, we obtained the reduced action of the matter
thin shell in AdS, and the stationary and stability conditions.  There
are two equations for the stationary condition, i.e., the balance of
pressure and the balance of temperature, and two stability conditions,
i.e., the mechanical stability condition and the thermodynamic
stability condition. The fact that one obtains the two stability
conditions shows the power of the reduced action in the Euclidean path
integral approach, in that it gives not only information about
thermodynamics but also of mechanics.

We showed that for the case of the matter thin shell in AdS, one can
obtain the canonical ensemble of the thin shell by establishing an
effective reduced action only dependent on the gravitational radius of
the thin shell. This eases the analysis of the
canonical
ensemble and
further shows that one can build an effective reduced action dependent
only on the gravitational radius for this case, hiding the description
of matter in the form of the effective entropy. Here, we established
the link between the effective entropy and the specific description of
the shell with an equation of state, in the zero-loop approximation.

The thermodynamics of the system follows directly from zero-loop
approximation consisting of the reduced action evaluated at the
stationary points, which is equivalent to finding the action for the
specific solution of Einstein equation.  In this approximation, one
obtains directly the relevant thermodynamic quantities, namely, the
mean free energy, the entropy, the mean energy, and the heat capacity.
There is a correspondence between thermodynamic stability in the
ensemble theory and positive heat capacity in the derived
thermodynamics, as it should.  On the other hand, within
thermodynamics itself one cannot determine mechanical stability by
varying the thermodynamic quantity fixed at the conformal boundary,
i.e., the temperature. This fact seems to be a consequence of applying
the zero-loop approximation to the internal degree of freedom, the
radius of the shell. It also means that the zero-loop approximation of
the effective reduced action yields the expected thermodynamic
stability condition, since it can be seen as a generalized free energy
function.

We introduced an equation of state and obtained the solutions of the
canonical ensemble for the matter thin shell. There are in total four
solutions, with only one of them being stable both mechanically and
thermodynamically. We also have compared the action of the stable
self-gravitating
matter shell solution with the Hawking-Page AdS black hole stable
solution and verified the existence of a first order phase transition
in a physically reasonable range of scale lengths.  We showed that
this first order phase transition is analogous to
the Hawking-Page phase transition, and that 
the hot matter thin shell can mimic nonself-gravitating radiation
in AdS. 
We note also that the phase structure is dependent on the 
choice of the equation of state and further study of this configuration 
for different families of equations of state should be done to uncover
other possible phase structures.
It will also be interesting to uncover the sector with a black hole and a
shell together to fully understand the space of configurations and
respective phase transitions. 

\section*{\vskip -1.0cm Acknowledgements}

The authors thank financial support from Funda\c c\~ao para a Ci\^encia 
e Tecnologia-FCT, Portugal, through the Project No.~UID/PRR/00099/2025 and
Project No.~UID/00099/2025. T.~F. thanks financial 
support through the Grant FCT No.~RD1415.

\section*{Data Availability}

The data that support the findings of this article are openly 
available in \cite{dataavailability}.

\end{document}